\documentclass[%
 print,
 amsmath,amssymb,
 aps,
]{revtex4-2}

\usepackage{graphicx}
\usepackage{dcolumn}
\usepackage{bm}
\usepackage{subfigure}
\usepackage{hyperref}

\usepackage{xcolor}
\hypersetup{
    colorlinks = true,
    citecolor  = blue,
    linkcolor  = blue,
    urlcolor   = blue,
}

\begin{document}

\preprint{APS/123-QED}

\title{First-step experiment in developing optical-spring quantum locking for DECIGO: sensitivity optimization for simulated quantum noise by completing the square}

\author{Tomohiro Ishikawa${}^1$}
\author{Yuki Kawasaki${}^1$}
\author{Kenji Tsuji${}^1$}
\author{Rika Yamada${}^1$}
\author{Izumi Watanabe${}^1$}
\author{Bin Wu${}^1$}
\author{Shoki Iwaguchi${}^1$}
\author{Ryuma Shimizu${}^1$}
\author{Kurumi Umemura${}^1$}
\author{Koji Nagano${}^2$}
\author{Yutaro Enomoto${}^3$}
\author{Kentaro Komori${}^4$}
\author{Yuta Michimura${}^{5,6}$}
\author{Akira Furusawa${}^{3,7}$}
\author{Seiji Kawamura${}^{1,8}$}

\affiliation{%
 ${}^1$Department of Physics, Nagoya University, Nagoya, Aichi 464-8602, Japan
}%
\affiliation{%
 ${}^2$Institute of Space and Astronautical Science, Japan Aerospace Exploration Agency, Sagamihara, Kanagawa 252-5210, Japan
}%
\affiliation{%
 ${}^3$Department of Applied Physics, School of Engineering, The University of Tokyo, 7-3-1 Hongo, Bunkyo-ku, Tokyo 113-8656, Japan
}%
\affiliation{%
 ${}^4$Department of Physics, University of Tokyo, Bunkyo, Tokyo 113-0033, Japan
}%
\affiliation{%
 ${}^5$LIGO Laboratory, California Institute of Technology, Pasadena, California 91125, USA
}%
\affiliation{%
 ${}^6$Research Center for the Early Universe (RESCEU), Graduate School of Science, University of Tokyo, Bunkyo, Tokyo 113-0033, Japan
}%
\affiliation{%
 ${}^7$Center for Quantum Computing, RIKEN, 2-1 Hirosawa, Wako, Saitama 351-0198, Japan
}%
\affiliation{%
 ${}^8$The Kobayashi-Maskawa Institute for the Origin of Particles and the Universe, Nagoya University, Nagoya, Aichi 464-8602, Japan
}%

\date{\today}

\begin{abstract}
DECi-hertz Interferometer Gravitational Wave Observatory (DECIGO) is a future mission for a space-borne laser interferometer. DECIGO has 1,000-km-long arm cavities mainly to detect the primordial gravitational waves (PGW) at lower frequencies around 0.1 Hz. Observations in the electromagnetic spectrum have lowered the bounds on the upper limit of PGW energy density ($\Omega_{\rm gw} \sim 10^{-15}\to10^{-16}$). As a result, DECIGO's target sensitivity, which is mainly limited by quantum noise, needs further improvement. To maximize the feasibility of detection while constrained by DECIGO's large diffraction loss, a quantum locking technique with an optical spring was theoretically proposed to improve the signal-to-noise ratio of the PGW. In this paper, we experimentally verify one key element of the optical-spring quantum locking: sensitivity optimization by completing the square of multiple detector outputs. This experiment is operated on a simplified tabletop optical setup with classical noise simulating quantum noise. We succeed in getting the best of the sensitivities with two different laser powers by the square completion method.
\end{abstract}

\maketitle


\section{Introduction}
\label{sec:1}

Over 90 gravitational-wave (GW) events have been detected \cite{Abbott021053,Abbott03606} by the international network (LIGO \cite{Aasi074001}, Virgo \cite{Acernese024001}, and KAGRA \cite{Somiya124007}). The detectors' designs are being updated to improve these sensitivities for the next observation run. The ground-based GW detectors have displacement noise such as seismic noise from the earth, gravity gradient noise, and thermal noise from mirror suspension at lower frequencies below 100~Hz. Some of these new designs aim to reduce or to cancel the displacement noise at the low frequencies. Einstein Telescope \cite{Punturo084007} and Cosmic Explorer \cite{Dwyer082001} are designed as the ground-based detectors for the next generation. A juggled interferometer \cite{Bin042007} and a (neutron) displacement noise-free interferometer \cite{Kawamura211103,Iwaguchi128150} have unique ideas to treat some displacement noise, and can be used on the ground. Plans for future GW detectors include space borne operations such as LISA \cite{Danzmann20170120} and BBO \cite{Crowder083005}.

DECi-hertz Interferometer Gravitational Wave Observatory (DECIGO) \cite{Seto221103,Kawamura1845001} is one of these space-borne interferometers. DECIGO will be operated mainly to detect the primordial GWs (PGW) \cite{Maggiore283367}, which is thought to have been produced during the inflation period. DECIGO's target sensitivity, which is mainly limited by quantum noise \cite{Caves3068,Schumaker3093}, was originally determined to be capable of detecting the PGW \cite{Seto221103}. However, observations of the Planck satellite \cite{Akrami641A10} and other electromagnetic observations \cite{Kuroyanagi063513} have lowered the upper limit of the normalized GW energy density for the PGW to $\Omega_\mathrm{gw}\sim10^{-16}$. Thus, we are working on further improving the target sensitivity of DECIGO to maximize the feasibility of detecting the PGW. A signal-to-noise ratio (SNR) to the PGW is maximized by optimizing DECIGO's design parameters while considering practical constraints \cite{Iwaguchi9010009,Ishikawa9010014,Kawasaki10010025}. Other SNR optimization methods aim at applying optical techniques such as employing auxiliary arm-scale cavities for input filtering \cite{Kimble022002} and injecting quantum squeezing into DECIGO's arm cavities \cite{LIGOCollab2083,Aasi177,Acernese321108}. Unfortunately, these ideas are not effective because large optical diffraction loss from the 1,000-km-long cavities \cite{Safia003805} lowers the enhancement from squeezing. Heavy mirrors cannot be installed in DECIGO's cavities because of the limited maximum loading weight for a spacecraft. A quantum locking technique was proposed to overcome the barriers to improving the target sensitivity \cite{Yamada126626,Yamada127365}.

\begin{figure}
    \centering
      \includegraphics[keepaspectratio,clip,width=12.0cm]{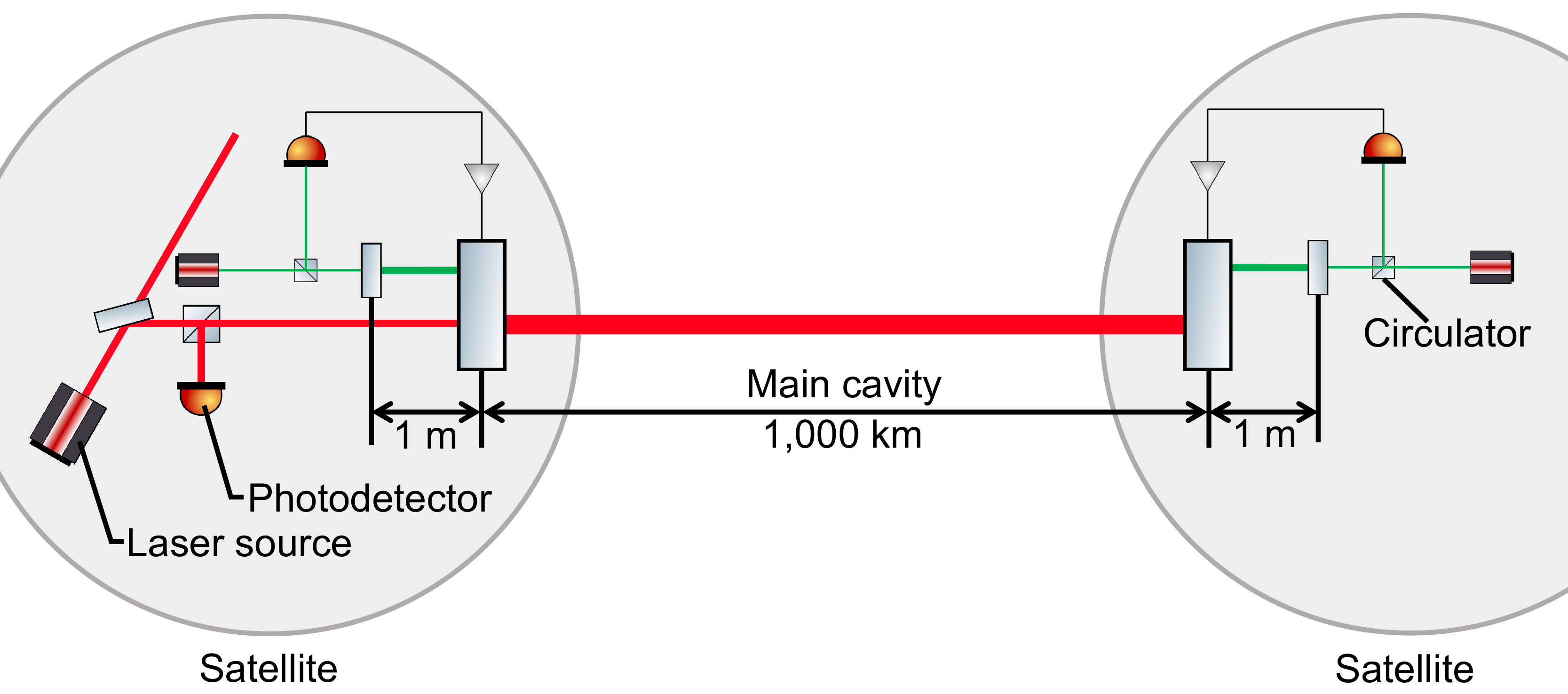} 
    \caption{Schematic geometry of a single arm in DECIGO with quantum locking. Each mirror in the arm cavity (red) is shared with a sub-cavity (green) located inside a satellite. The output signal from each sub-cavity is individually fed back to the respective shared mirror.}
    \label{fig:1}
  \end{figure}

Quantum locking is a mirror control technique for the arm cavity using an auxiliary short cavity \cite{Courty083601,HeidmannS684}. Figure \ref{fig:1} shows a simplified layout of an arm cavity in DECIGO with the quantum locking technique; the two mirrors of the long arm cavity (main cavity) are shared by short low-loss cavities (sub-cavities). The motion of the shared mirrors is controlled by feedback signals from the sub-cavities. A sensitivity optimization method and a homodyne detection are added to the quantum locking \cite{Yamada126626}. The former optimizes a target sensitivity by combining signals of the main cavity and the two sub-cavities. Introducing the homodyne detection to the sub-cavities allows for sensitivity improvement beyond the standard quantum limit (SQL). An optical spring effect \cite{Arcizet033819,Verlot133602} is also added in the sub-cavities to the quantum locking \cite{Yamada127365}. The effect widens the dip in the sensitivity curve made by the homodyne detection. It was theoretically shown that the quantum locking with the optical spring for DECIGO could drastically improve the SNR \cite{Yamada127365}. Therefore, demonstrating the optical-spring quantum locking in an experiment is important for DECIGO to maximize the sensitivity to the PGW.

We do an experiment to individually verify key elements in the optical-spring quantum locking (the sensitivity optimization method, the homodyne detection, and the optical spring effect). In this paper, we focus on the sensitivity optimization method, which is valid even without the homodyne detection and the cavity detuning for the optical spring. This fact enables us to fix the readout quadrature of the laser light to a phase quadrature in the experiment. Thus, we experimentally verified the method in a simplified, tabletop configuration and with a classical noise that simulates the quantum noise.

This paper is organized as follows. We review the optimization method in Sec.~\ref{sec:2}. Sec.~\ref{sec:3} details the experimental setup. We present and discuss results in Sec.~\ref{sec:4}, and conclude the research in Sec.~\ref{sec:5}.

\section{Theory}
\label{sec:2}

\begin{figure}[htbp]
    \centering
      \includegraphics[keepaspectratio,clip,width=8.0cm]{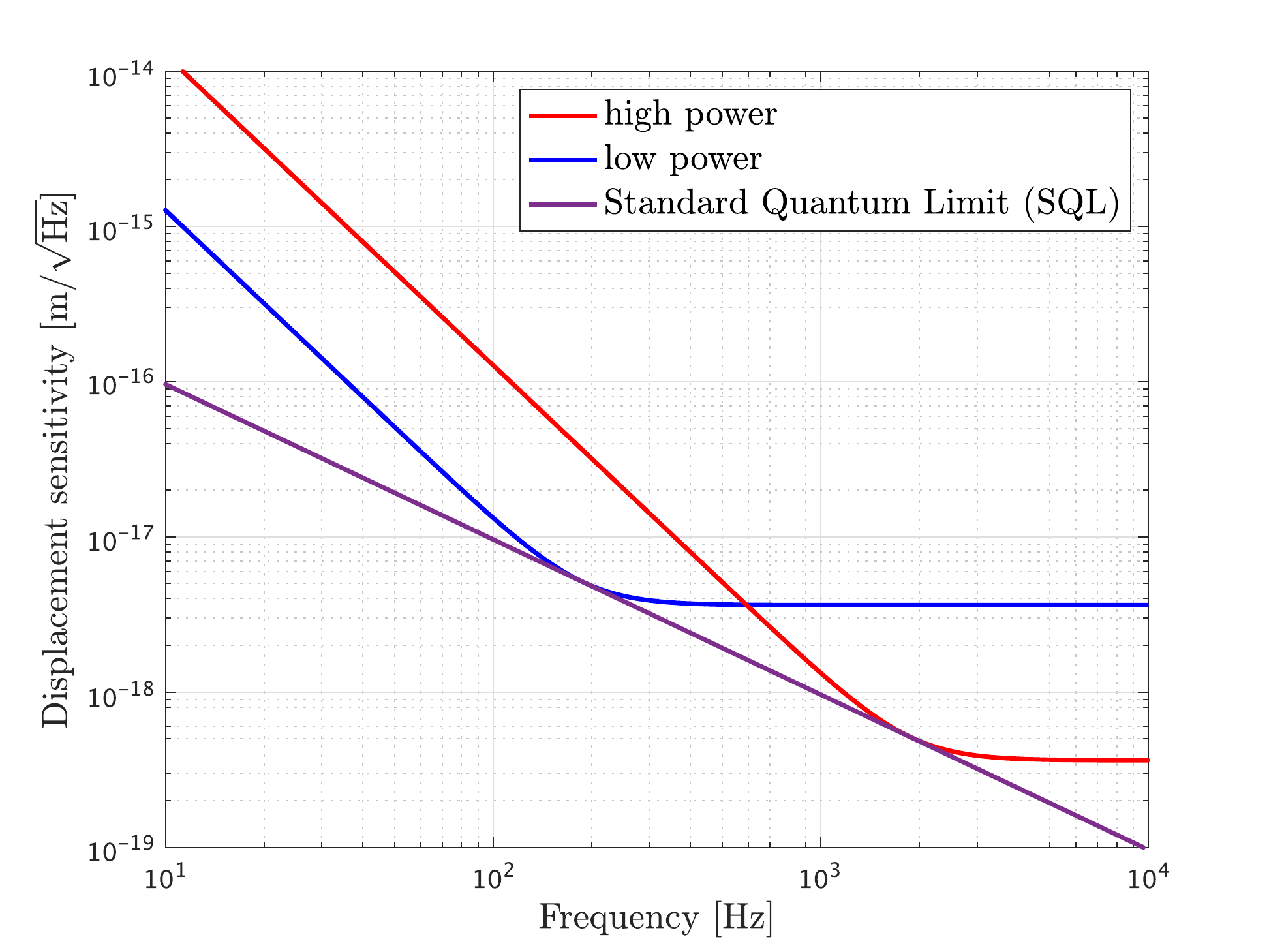}
    \caption{Quantum noise sensitivities for two different laser powers. The quantum noise contains a trade-off relation between the shot noise and the RP noise. The tangent line to the two quantum-noise curves shows the SQL.}
    \label{fig:2}
  \end{figure}

The quantum noise consists of two components: shot noise and radiation pressure (RP) noise \cite{Caves75,Caves1693}. The shot noise, independent of frequency, is due to the photon counting error at a photodetector (PD). The RP noise arises from fluctuations of the laser power, which when optomechanically coupled to the mirror mass creates a frequency dependent displacement noise. Figure \ref{fig:2} illustrates quantum-noise-limited sensitivities of a single arm cavity with two different laser powers $P_{\rm High}$ and $P_{\rm Low}$. The figure shows that we can obtain a better sensitivity at higher/lower frequencies when the laser power is higher/lower. It would be desirable if we could select better sensitivities depending on frequencies (cherry picking). However, selection of the sensitivities is impossible because the two sensitivities cannot be achieved simultaneously in a single cavity with a fixed circulating power.

Fortunately, we can do this ``cherry picking'' with the help of the quantum locking technique. Let us consider the sensitivity of the main arm cavity in Fig.~\ref{fig:1}. We compare the sensitivity in two cases when a servo loop gain is infinitely large (with the servo) and the gain is zero (without the servo). We assume that the laser power of the main/sub- cavity ($P_{\rm Main},~P_{\rm Sub}$) is related to $P_{\rm High}$ or $P_{\rm Low}$: $P_{\rm Main} = P_{\rm High}$, $P_{\rm Sub} = P_{\rm Low}/2$. The latter relation adjusts the magnitude of the quantum noise in the main cavity derived from the sub-cavities to that of the main cavity with $P_{\rm Low}$. We also assume that all the cavities are on resonance, all the mirrors in the cavities have no optical loss, and the cavity pole frequency is much higher than the relevant frequencies.

In the case without the servo, the sensitivity of the main cavity is the same as that without the auxiliary cavities, except that the RP noise is only slightly larger; due to the smaller RP noise in the sub-cavities being added to the RP noise of the main cavity via optomechanical coupling to the shared mirrors. In the case with the servo, the sensitivity is modified by the quantum noise of the sub-cavities. The feedback signals suppress the motion of the shared mirrors by the larger RP noise from $P_{\rm Main}$ but add the smaller RP noise from $P_{\rm Sub}$. The signals also add the larger shot noise from $P_{\rm Sub}$. This implies that the presence or absence of the feedback signals switches the sensitivity of the main cavity between the higher-power and the lower-power cases in Fig.~\ref{fig:2}. Then, the question is how we can obtain the best of both sensitivities over the entire frequency band.

One of the solutions to obtain the ``cherry-picked'' sensitivity is to optimize the quantum locking system's loop gain for the sensitivity. However, this solution could make stable control of the cavities challenging. Therefore, another solution is proposed \cite{Yamada126626}: to optimize a combined signal from output signals of the main cavity and the sub-cavities.

\begin{figure}[htbp]
    \subfigure[]{%
      \centering
        \includegraphics[keepaspectratio,clip,width=8.0cm]{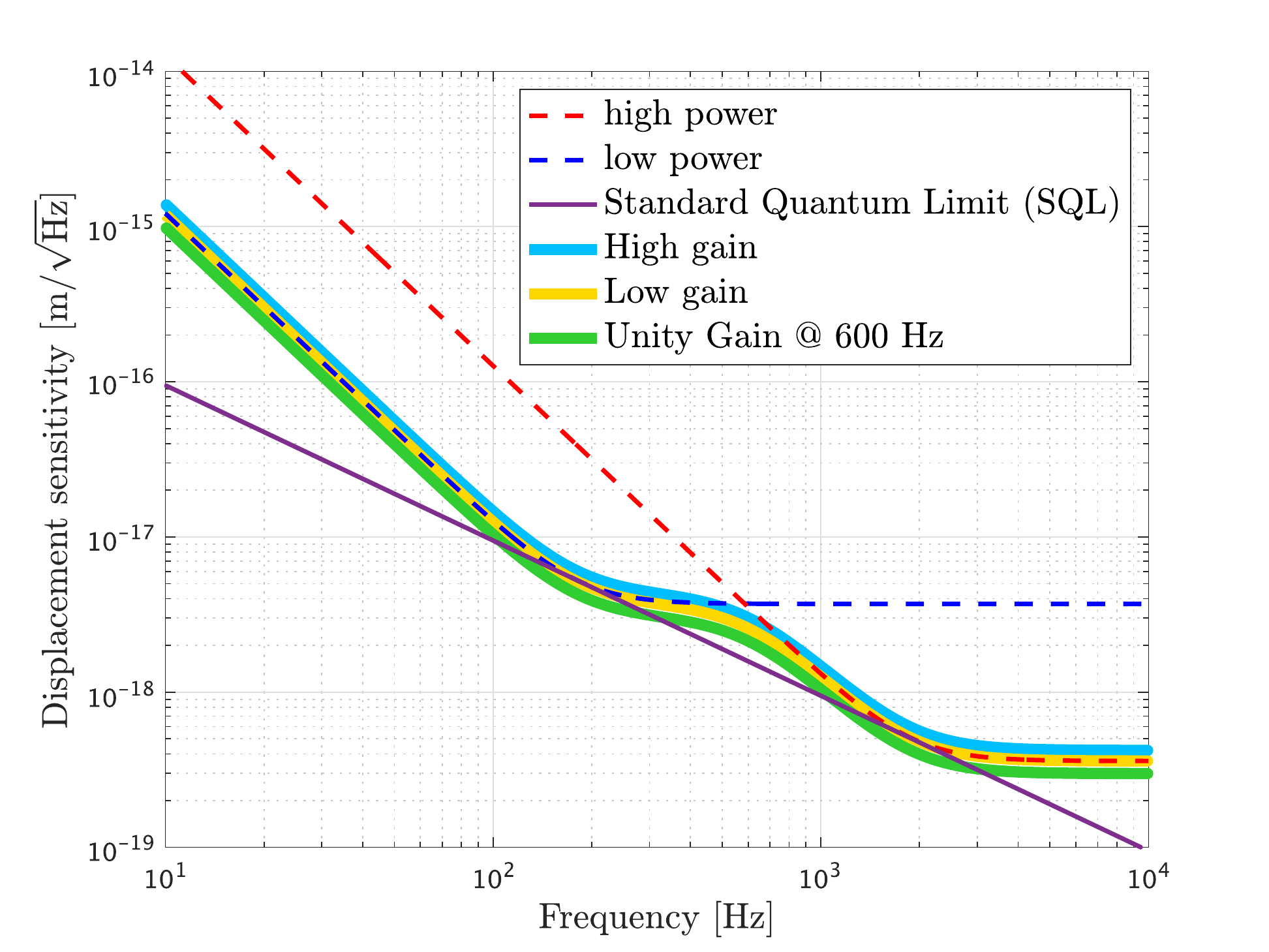}
    } %
    \subfigure[]{%
      \centering
        \includegraphics[keepaspectratio,clip,width=8.0cm]{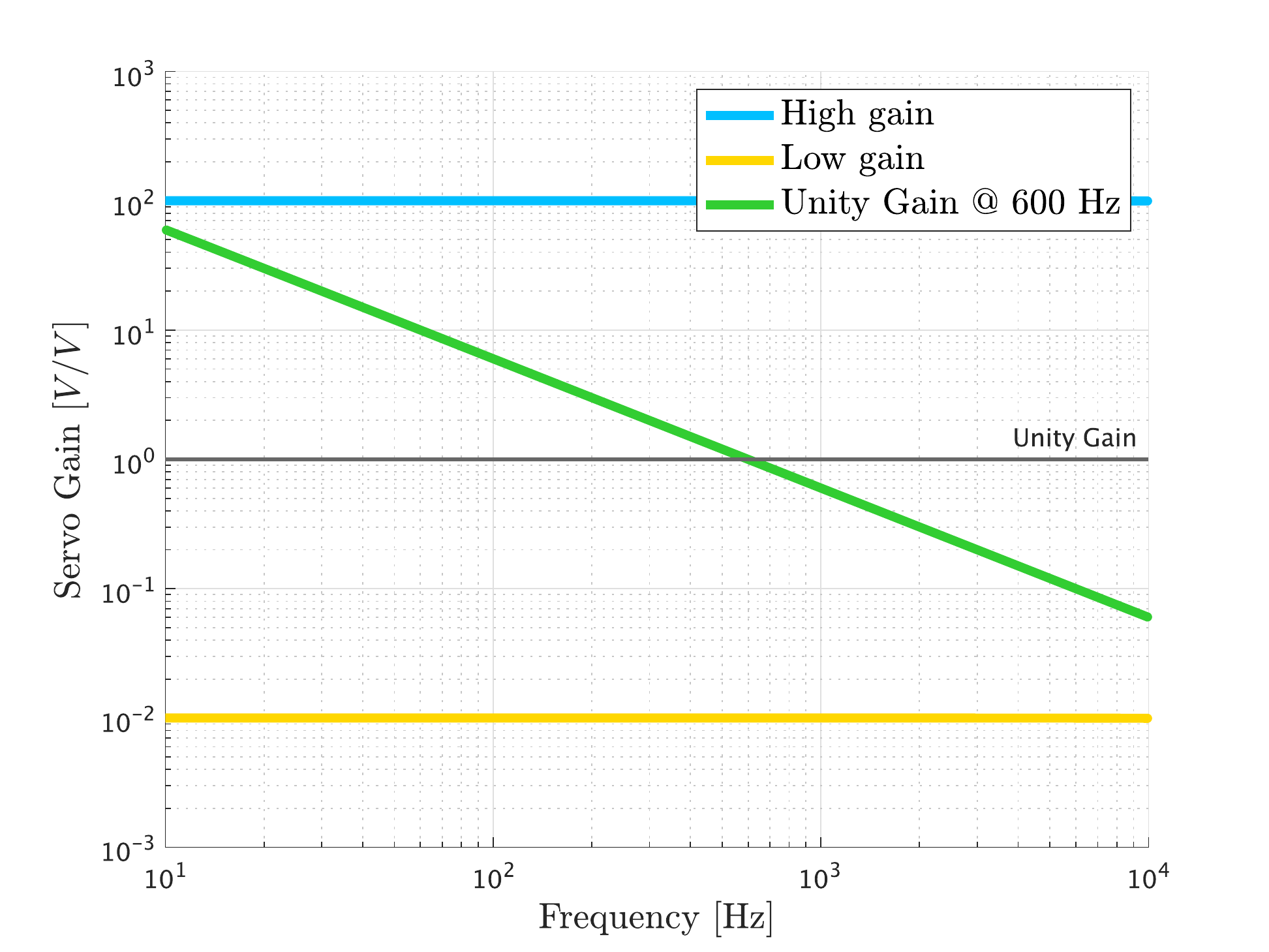}
    }%
    \caption{Simulated sensitivities for the optimal combined signal $\tilde{V}_\mathrm{opt}$ in the three cases: high, low, and typical loop gain. (a) includes the optimized sensitivities and two sensitivities of the single arm cavity with high/low laser power. The applied loop gain is shown in (b). For better visibility of the optimized sensitivities in the plot (a), we slightly displace the blue trace upward, and the green trace downward from the yellow curve.}
    \label{fig:3}
  \end{figure}

We define the combined signal $\tilde{V}$ with an arbitrary complex function $\chi$ as
  \begin{equation}
      \tilde{V}(f) \equiv \tilde{V}_{\rm Main} (f) + \chi(f) \left( \tilde{V}_{\rm Sub1} (f) + \tilde{V}_{\rm Sub2} (f) \right)~.
    \label{eq:2-1}
  \end{equation}
Here $\tilde{V}_{\rm Main}$ is a Fourier transformed output signal of the main cavity, and $\tilde{V}_{{\rm Sub}i}~(i=1,2)$ are those of the sub-cavities. The total noise of $\tilde{V}$ is a quadrature sum of all the noise components in Eq.~\eqref{eq:2-1}. Here, we use a square completion method for $\tilde{V}$, which derives a minimum value of an arbitrary quadratic function at every frequency. The method can provide an optimized signal $\tilde{V}_\mathrm{opt}$ by an appropriate function $\chi_\mathrm{opt}$. Figure \ref{fig:3} plots the optimized sensitivity of $\tilde{V}$ by completing the square for the three cases: high, low, and typical loop gain. The figure shows that the sensitivity is optimized by the method to a unique value which does not depend on the applied loop gain. This favorable result is based on the fact that the proper coefficient $\chi_{\rm opt}$ is a function of the loop gain.

Figure \ref{fig:3} also shows that, at the cross-over frequency between the two sensitivities of a single cavity, the optimized sensitivity is enhanced by a factor of $1/\sqrt{2}$. We intuitively explain the mechanism for a simple case without the servo. As shown in Fig.~\ref{fig:3}, the main/sub- cavity is dominated by the RP/shot noise from $P_{\rm Main}$ or $P_{\rm Sub}$ at the cross-over frequency. From the geometry in Fig.~\ref{fig:1} and the zero loop gain, the signals $\tilde{V}_{\rm Main}$ and $\tilde{V}_{{\rm Sub}i}$ are related to the GW signal $s$, the RP noise from $P_{\rm Main}$, $\tilde{V}^{\rm RP}_{\rm Main}$, and the shot noise from $P_{\rm Sub}$, $\tilde{V}^{\rm Shot}_{{\rm Sub}i}$ as:
  \begin{equation}
      \tilde{V}_{\rm Main} = s + \tilde{V}^{\rm RP}_{\rm Main},~\tilde{V}_{{\rm Sub}i} = - \frac{1}{2} \tilde{V}^{\rm RP}_{\rm Main} + \tilde{V}^{\rm Shot}_{{\rm Sub}i}~.
    \label{eq:2-2}
  \end{equation}
All the noise components are independent. Note that these power spectral densities have a relation, $S^{\rm RP}_{\rm Main} = S^{\rm Shot}_{{\rm Sub}i}/2$, which comes from the relation of the laser power. Equation \eqref{eq:2-2} follows that the combined signal $\tilde{V}$ is minimized when $\chi = 1/2$. The optimized signal $\tilde{V}_{\rm opt}$ is calculated by Eqs.~\eqref{eq:2-1} and \eqref{eq:2-2}:
  \begin{equation}
      \tilde{V}_{\rm opt} = s + \frac{1}{2} \left( \tilde{V}^{\rm RP}_{\rm Main} + \tilde{V}^{\rm Shot}_{\rm Sub} \right)
    \label{eq:2-3}
  \end{equation}
where $\tilde{V}^{\rm Shot}_{\rm Sub}$, whose power spectral density is the same as that of $\tilde{V}^{\rm RP}_{\rm Main}$, is a quadrature sum of the shot noise $\tilde{V}^{\rm Shot}_{{\rm Sub}i}$. Therefore, the optimized sensitivity at the cross-over frequency is improved by a factor of $\sqrt{2}/2 = 1/\sqrt{2}$ when compared to the single cavity case.

\section{Experiment}
\label{sec:3}

\subsection{Breaking down from theory to experiment}
\label{sec:3-1}

As mentioned in Sec.~\ref{sec:1}, we fixed the readout quadrature of the laser light to the phase quadrature in this experiment. The phase quadrature of the light directly appears as the shot noise at the interferometer output, while the amplitude quadrature indirectly appears as the RP noise at the output via the mirror motion. Therefore, all we need are two independent noise components: one at the output for the shot noise and the other at the mirror displacement for the RP noise. We can add the classical noise from a signal generator, which simulates the quantum noise, directly to the output signal for the shot noise and to the mirror motion for the RP noise. We call this classical noise a modeled quantum noise (more specifically, a modeled shot noise and a modeled RP noise).

\begin{figure*}[htbp]
    \centering
      \includegraphics[keepaspectratio,clip,width=12.5cm]{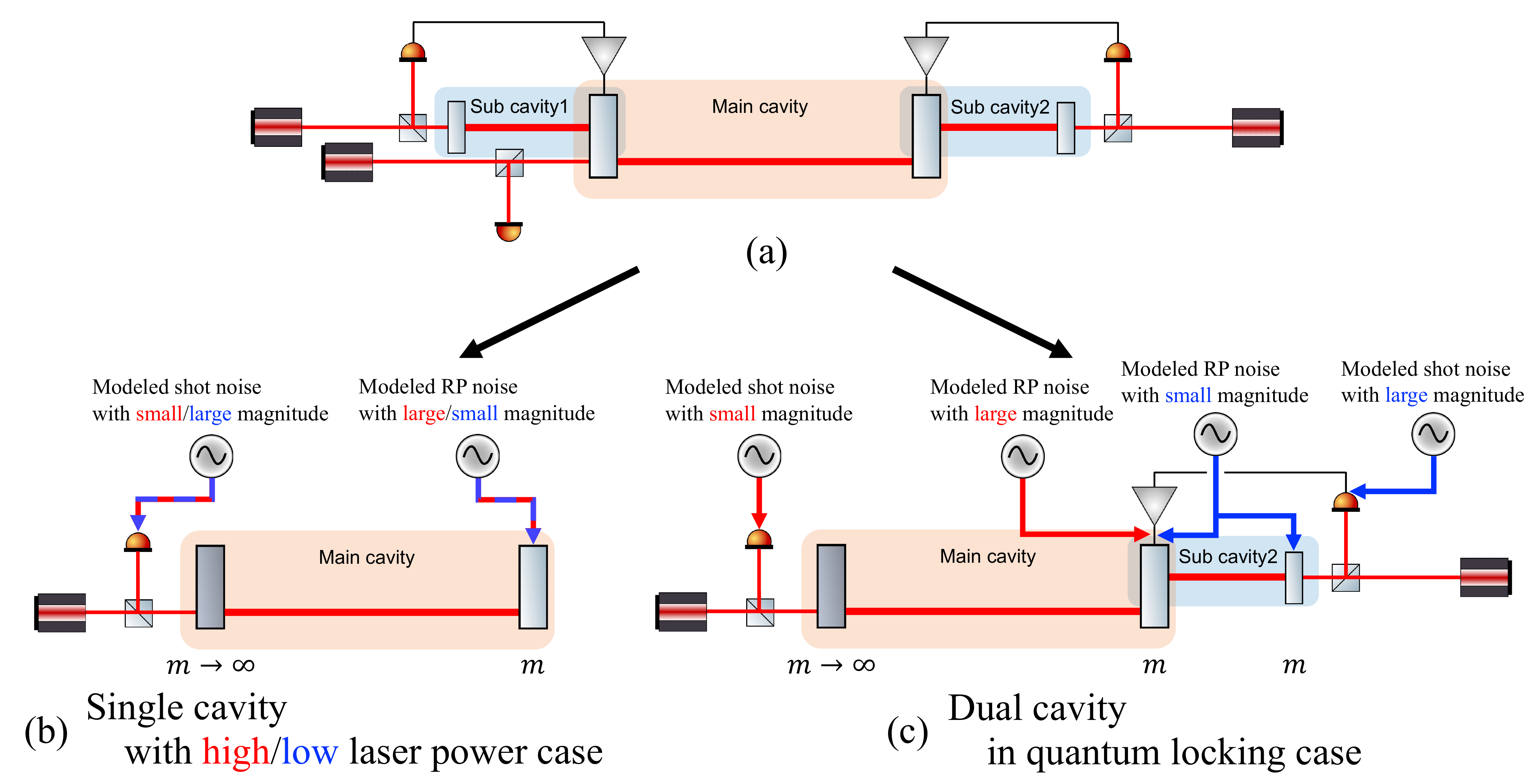}
    \caption{Simplified layout with the modeled quantum noise for the three cases. The essence of the optical setup (a), the same as Fig.~\ref{fig:1}, is broken down into two cases; (b) the single cavity case with the high/low laser power and (c) the dual cavity case. A mirror, which is assumed to have infinite mass, is colored dark gray. Arrows from the signal generators show the injection points.}
    \label{fig:4}
  \end{figure*}

We specify the essence of the principle to simplify the optical layout in Fig.~\ref{fig:4}(a) mimicking the optical configuration in Fig.~\ref{fig:1}. From the discussion in Sec.~\ref{sec:2}, the combined signal in Eq.~\eqref{eq:2-1} is optimized by the square completion method. The sensitivity of the optimal combined signal follows the sensitivity of the single cavity with the high/low laser power at the high/low frequency. Thus, we measure three modeled sensitivities: the two sensitivities of the single cavity with the modeled quantum noise corresponding to the high/low laser power and the optimal sensitivity from the combined signal in Eq.~\eqref{eq:2-1}. We then compare the three sensitivities to verify that the latter optimal sensitivity truly reflects the best of the former two sensitivities. Note that the magnitude of the modeled quantum noise includes the influence of the laser power to the quantum noise, thereby the actual laser powers for the main/sub- cavity are arbitrarily assigned.

We consider the following setup in the experiment. We presume that one shared mirror of the main cavity in Fig.~\ref{fig:4}(a) is infinitely heavy. This allows us to simplify the setup by omitting the sub-cavity whose shared mirror has an infinite mass. Modeled quantum noises corresponding to the laser powers $P_{\rm Main}$ and $P_{\rm Sub}$ are the same as that of $P_{\rm High}$ and $P_{\rm Low}$, respectively. This implies that $P_{\rm Main} = P_{\rm High}$ and $P_{\rm Sub} = P_{\rm Low}$. Every shot noise component is directly added to appropriate output signals, while every RP noise component is added by shaking appropriate mirrors except for the mirrors assumed to have an infinite mass.

With the simplified setup, we digest the operations for the three cases. When we measure the two sensitivities of the single cavity, we can only operate the main cavity, as shown in Fig.~\ref{fig:4}(b). While measuring the optimal sensitivity, we operate both cavities, as shown in Fig.~\ref{fig:4}(c). Note that the two mirrors of the sub-cavity are shaken in a differential phase by the modeled RP noise from $P_{\rm Sub}$ for consistency with the real quantum RP noise. We call the case of operating both cavities the dual cavity case. In the next subsection, we introduce a detailed experimental setup to verify the principle.

\newpage
\subsection{Setup and measurement}
\label{sec:3-2}

\begin{figure*}[htpb]
    \subfigure[]{%
      \centering
        \includegraphics[keepaspectratio,clip,width=8.0cm]{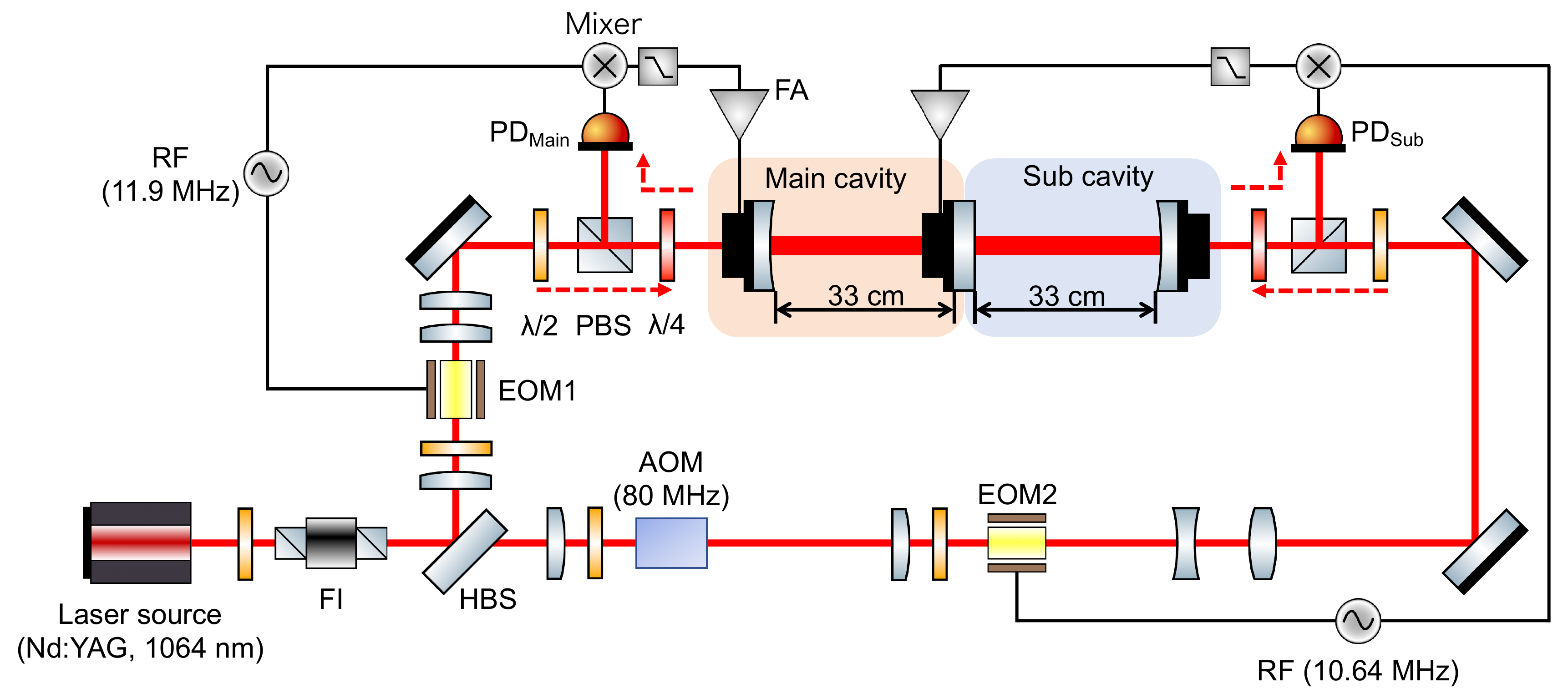}
      \label{fig:5-1}
    }%
    \subfigure[]{%
      \centering
        \includegraphics[keepaspectratio,clip,width=8.0cm]{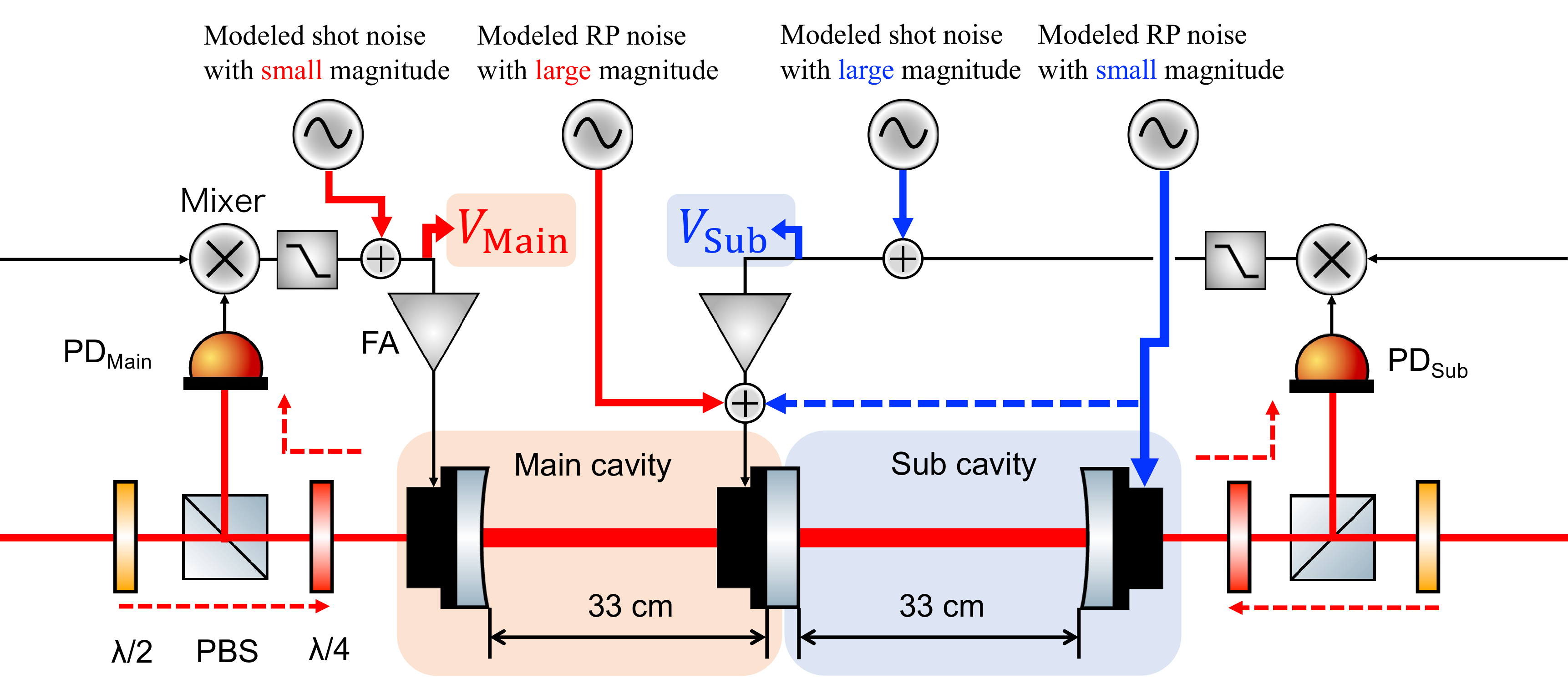}
      \label{fig:5-2}
    }%
    \caption{(a) Experimental setup overview, and (b) detailed control signal flow diagram for the two cavities with the classical noise injections and the measuring points. The two identically designed Fabry-Perot cavities share an end mirror and are individually controlled by separate feedback signals. The main/sub- cavity are individually operated in the high/low power single cavity case while both cavities are operated in the dual cavity case. The dashed blue arrow in (b) only exists in the dual cavity case. AOM: acousto-optic modulator, EOM: electro-optic modulator, FA: filter amplifier, FI: Faraday isolator, HBS: half beam splitter, PD: photodetector, PBS: polarizing beam splitter, RF: radio frequency; $\lambda/2$: half-wave plate, $\lambda/4$: quarter-wave plate.}
    \label{fig:5}
  \end{figure*}

Figure \ref{fig:5-1} contains an overview of the experimental setup. The main/sub- cavity are designed as two Fabry-Perot cavities with the same cavity length ($L = 33~\mathrm{cm}$) and the same finesse ($\mathcal{F} \sim 3\times10^2$). Both cavities have the same plane-concave configuration (g factor: $g \sim 0.3$). The input mirrors have a radius of curvature of $5\times10^2$~mm and the shared end mirror is flat. The end mirror is almost completely reflective ($R = 99.9~\%$). High reflectivity coatings are on both sides of the end mirror. The high-reflection surfaces are wedged to prevent interference between transmitted light from the two cavities. All the mirrors of the cavities are fixed on mirror holders via piezoelectric transducers (efficiency: 8~$\mu$m/150~V). We use the Pound-Drever-Hall (PDH) method \cite{Drever00702605} to sense the cavities' length. In Fig.~\ref{fig:5-1}, the cavity on the left side is the main cavity, and the cavity on the right side is the sub-cavity.

The 1,064~nm laser generated by a Nd:YAG laser source is divided into two paths by a half beam splitter. The two laser paths propagate over the same distance to each cavity. Phase modulation signals for the PDH method are individually applied to each path by electro-optic modulators. An acousto-optic modulator shifts the frequency of the laser light toward the sub-cavity by 80~MHz to avoid interference between the laser inside the two cavities. After passing through mode-matching lenses, the two paths enter their respective cavities. The light reflected from each cavity is detected by their respective PD. The error signal, which is obtained by mixing the PD signal by the phase modulation signal, is appropriately filtered and used as the feedback signal. The feedback signal of the main/sub- cavity is applied to the input/end mirror in the main/sub- cavity.

Following the discussion in Sec.~\ref{sec:3-1}, we need to measure/obtain the three sensitivity curves of the high/low power single cavity cases and the dual cavity case. We can operate straightforward settings for the high-power single cavity case and the dual cavity case. In the low-power case, it is possible for us to operate the sub-cavity instead of the main cavity as the single cavity because the cavities in this experiment are identically designed. The modeled quantum noise corresponding to the low laser power is applied to the sub-cavity in the dual cavity case. Thus we can operate the sub-cavity with the modeled quantum noise of the low laser power in the low-power case. This means that we operate the main/sub- cavity in the high/low power case and both cavities in the dual cavity case. At the sub-cavity, the modeled RP noise with the small magnitude is applied to both mirrors in the dual cavity case while it is applied to only the input mirror in the low-power case; the dashed blue arrow in Fig.~\ref{fig:5-2} means that the path exists only in the dual cavity case.

We individually detail the sensitivity calculations for the high/low power single cavity cases and in the dual cavity case. In the high/low power single cavity cases, we record a time-varying error signal $V_\mathrm{Main}(t)$ or $V_\mathrm{Sub}(t)$ at the measurement points in Fig.~\ref{fig:5-2}. The Fourier transformed error signal $\tilde{V}_\mathrm{Main}(f)$ or $\tilde{V}_\mathrm{Sub}(f)$ is calibrated in terms of displacement of the main/sub- cavity length. We then obtain the signal as displacement noise sensitivity. Note that the calibration function is computed from a simulation that reflects the experimental properties. The observation frequency range is set from 700~Hz to 10~kHz for stably controlling the cavities even with large and visible classical noise.

The optimal combination coefficient $\chi_\mathrm{opt}$ is separately prepared from $\tilde{V}_\mathrm{Main}$ and $\tilde{V}_\mathrm{Sub}$ in the dual cavity case. Every modeled quantum noise individually transfers from every noise source to each PD; we define transfer functions from every noise source to each PD as $A \equiv \tilde{V}_\mathrm{Main}/q_\mathrm{Main},~B \equiv \tilde{V}_\mathrm{Main}/p_\mathrm{Main},~C \equiv \tilde{V}_\mathrm{Main}/q_\mathrm{Sub},~D \equiv \tilde{V}_\mathrm{Main}/p_\mathrm{Sub},~E \equiv \tilde{V}_\mathrm{Sub}/q_\mathrm{Main},~F \equiv \tilde{V}_\mathrm{Sub}/p_\mathrm{Main},~G \equiv \tilde{V}_\mathrm{Sub}/q_\mathrm{Sub}$, and $H \equiv \tilde{V}_\mathrm{Sub}/p_\mathrm{Sub}$. Here $q_j,~p_j~(j=\mathrm{Main,~Sub})$ are magnitude-normalized modeled RP noise and modeled shot noise, respectively. Error signals $\tilde{V}_\mathrm{Main}$ and $\tilde{V}_\mathrm{Sub}$ can theoretically be described by $A \sim H$:
  \begin{align}
      \tilde{V}_\mathrm{Main}(f) &= A(f) q_\mathrm{Main} + B(f) p_\mathrm{Main} + C(f) q_\mathrm{Sub} + D(f) p_\mathrm{Sub}~, \label{eq:3-1} \\
      \tilde{V}_\mathrm{Sub}(f) &= E(f) q_\mathrm{Main} + F(f) p_\mathrm{Main} + G(f) q_\mathrm{Sub} + H(f) p_\mathrm{Sub}~.
    \label{eq:3-2}
  \end{align}
$A \sim H$ produce the optimal combination function $\chi_\mathrm{opt}$ by completing the square for the combined signal $\tilde{V} = \tilde{V}_\mathrm{Main} + \chi \tilde{V}_\mathrm{Sub}$ following Eq.~\eqref{eq:2-1}:
  \begin{equation}
      \chi_\mathrm{opt} (f) = - \frac{A E^* + B F^* + C G^* + D H^*}{|E|^2 + |F|^2 + |G|^2 + |H|^2}~.
    \label{eq:3-3}
  \end{equation}
To determine $\chi_\mathrm{opt}(f)$, the transfer functions $A \sim H$ calculated in the simulation with the experimental parameters are used in the experiment to obtain the optimal sensitivity.

We measure $V_\mathrm{Main}(t)$ and $V_\mathrm{Sub}(t)$ in the dual cavity case. We then calculate the optimal combined signal $\tilde{V}_\mathrm{opt}(f)$ by combining the measured signals $\tilde{V}_\mathrm{Main}(f)$ and $\tilde{V}_\mathrm{Sub}(f)$ with the simulated optimal coefficient $\chi_\mathrm{opt}$ in Eq.~\eqref{eq:3-3}:
  \begin{equation}
      \tilde{V}_\mathrm{opt} (f) = \tilde{V}_\mathrm{Main}(f) + \chi_\mathrm{opt} (f) \tilde{V}_\mathrm{Sub}(f)~.
    \label{eq:3-4}
  \end{equation}
The optimized signal $\tilde{V}_\mathrm{opt}$ is calibrated in terms of displacement of the main cavity's length in the same way as the single cavity case. This lets us measure the displacement noise sensitivity in the dual cavity case.

\begin{figure}[htbp]
    \subfigure[For the main cavity]{%
      \centering
        \includegraphics[keepaspectratio,clip,width=8.0cm]{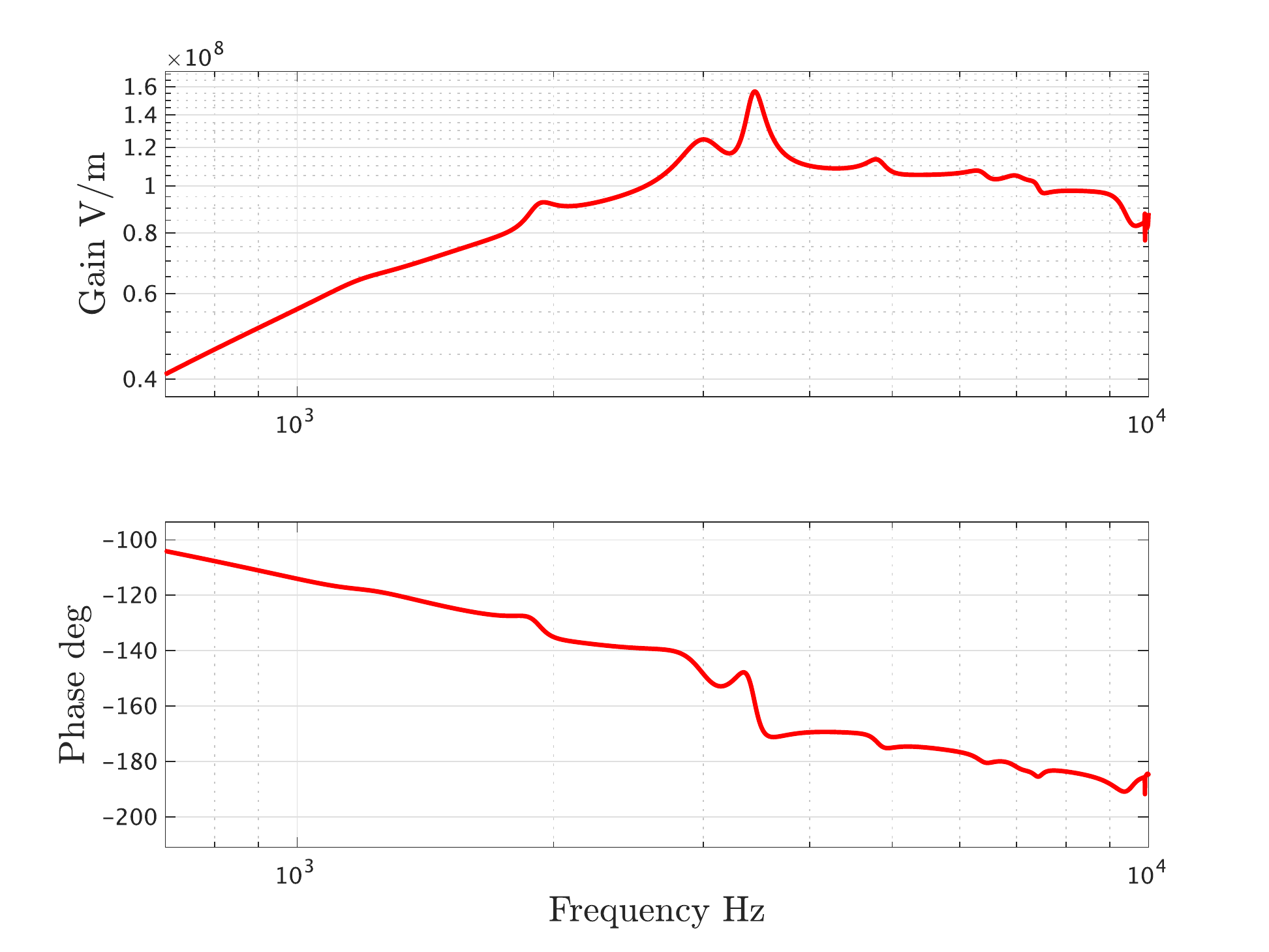}
      \label{fig:6-1}
    } %
    \subfigure[For the sub-cavity]{%
      \centering
        \includegraphics[keepaspectratio,clip,width=8.0cm]{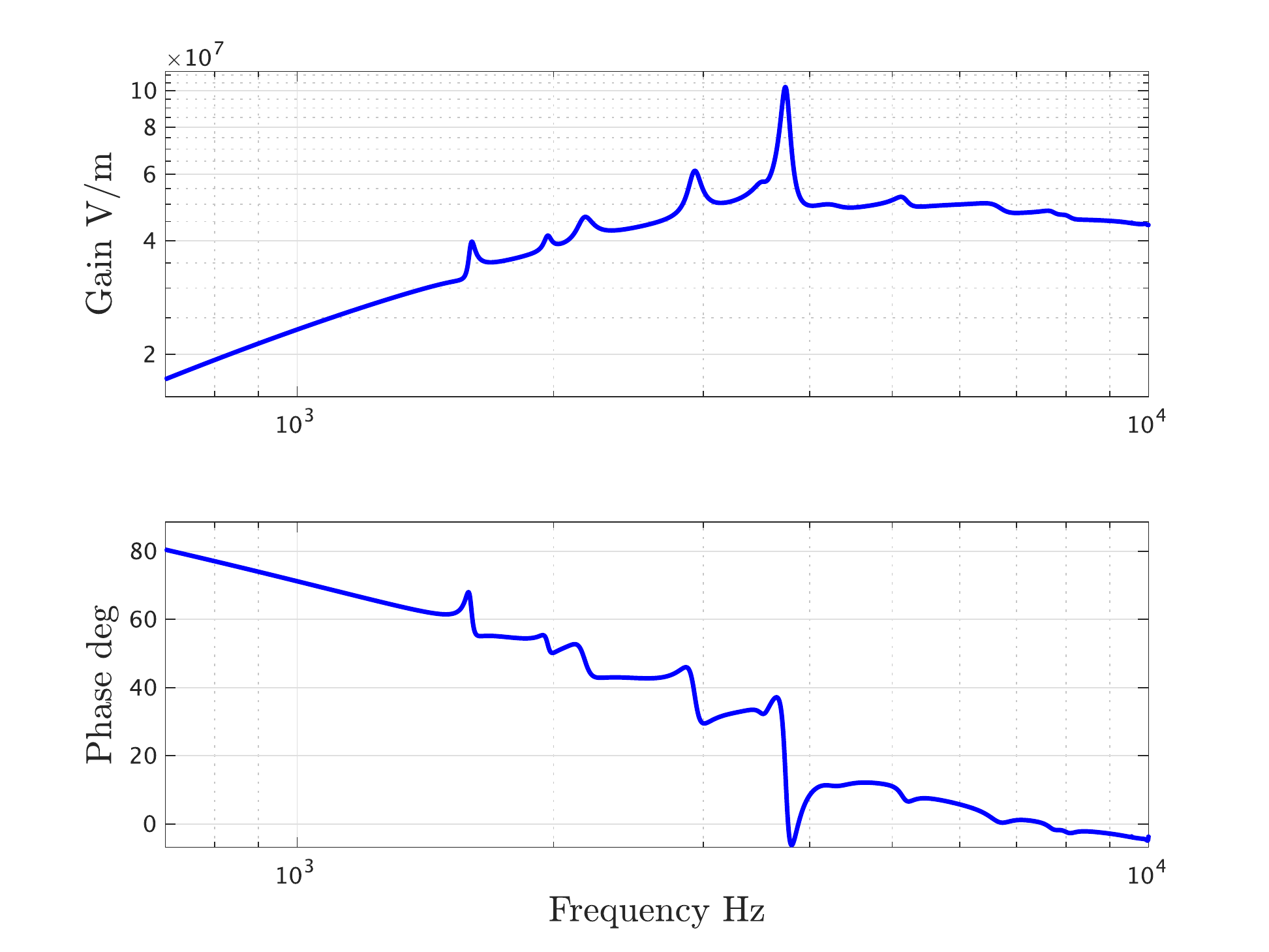}
      \label{fig:6-2}
    }%
    \caption{Corrected calibration functions at the observation frequency range from 700~Hz to 10~kHz. The functions for the main cavity (a) and for the sub-cavity (b) depict similar trends with different mechanical resonances.}
    \label{fig:6}
  \end{figure}

We should note that there are resonances from the mechanical structures holding the mirrors. We account for this influence by correcting all the simulated functions in the observation frequency range from 700~Hz to 10~kHz with the fitted frequency dependence of the resonances. Figure~\ref{fig:6} shows the frequency dependence of the corrected calibration functions.

\section{Result and Discussion}
\label{sec:4}

\begin{figure*}[htbp]
    \subfigure[]{%
      \hspace{5.7\baselineskip}
        \includegraphics[keepaspectratio,clip,width=11.0cm]{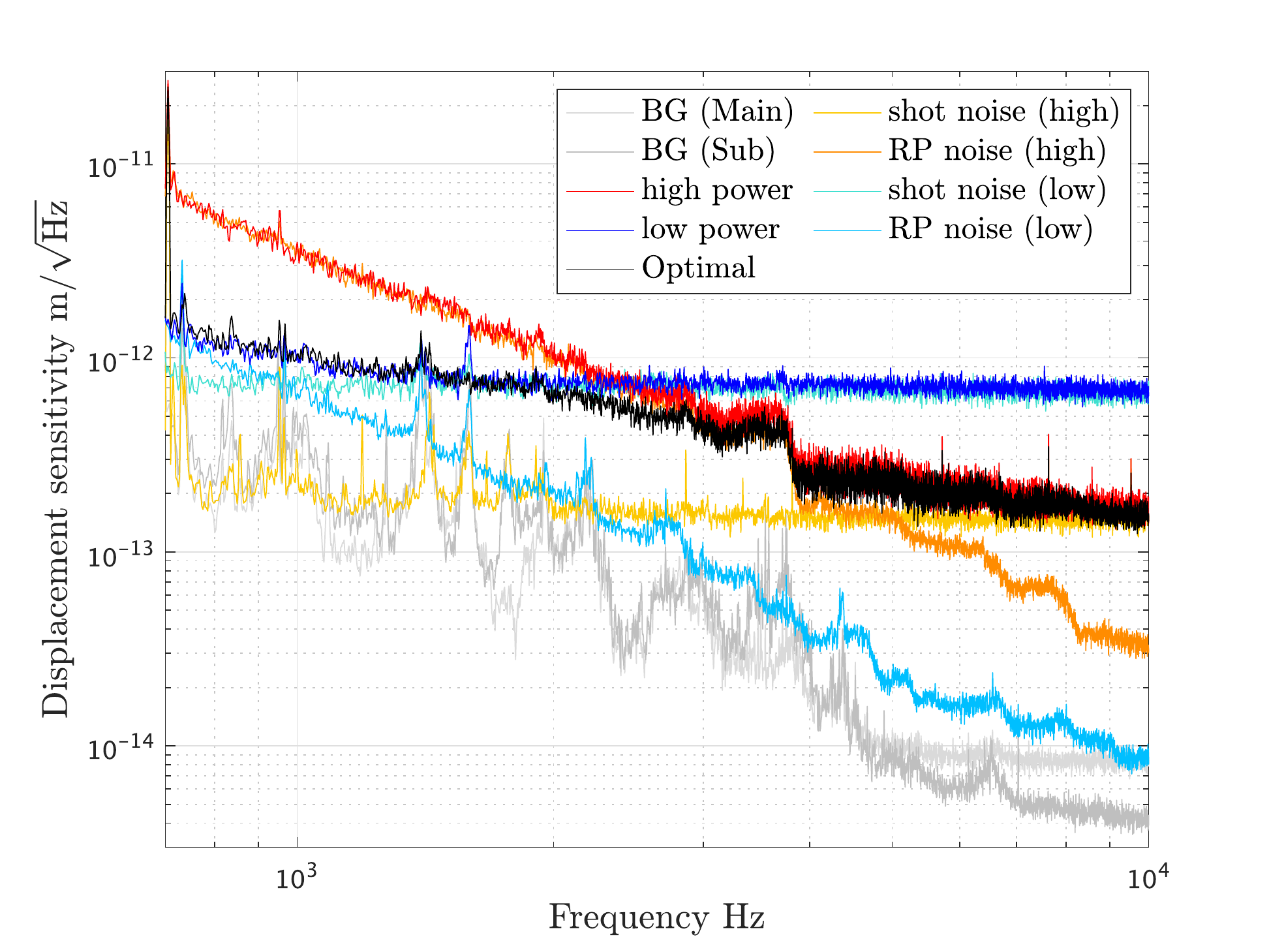}
      \label{fig:7-1} \hspace{5.7\baselineskip}
    } \\ %
    \subfigure[]{%
      \centering
        \includegraphics[keepaspectratio,clip,width=8.0cm]{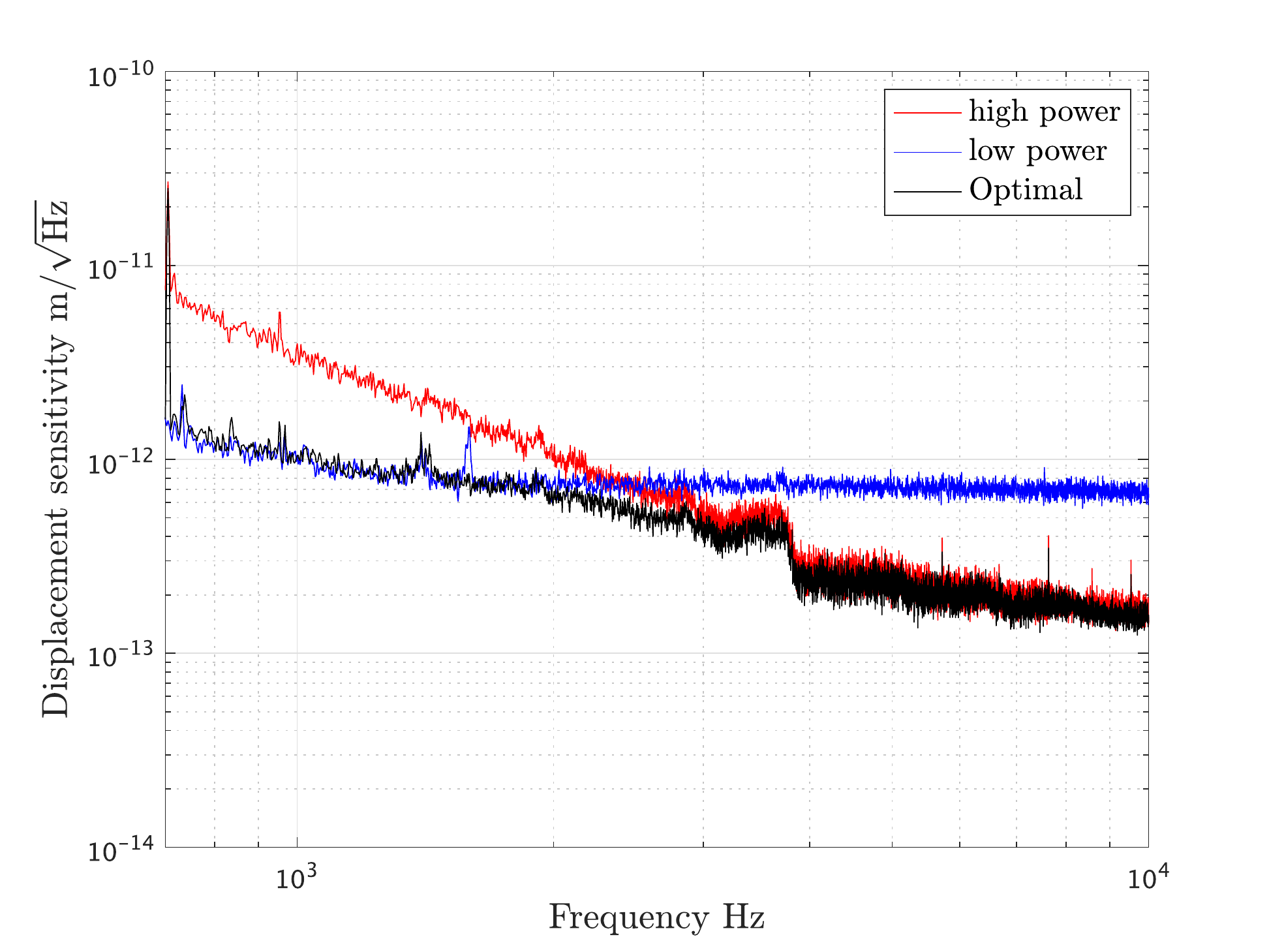}
      \label{fig:7-2}
    }%
    \subfigure[]{%
      \centering
        \includegraphics[keepaspectratio,clip,width=8.0cm]{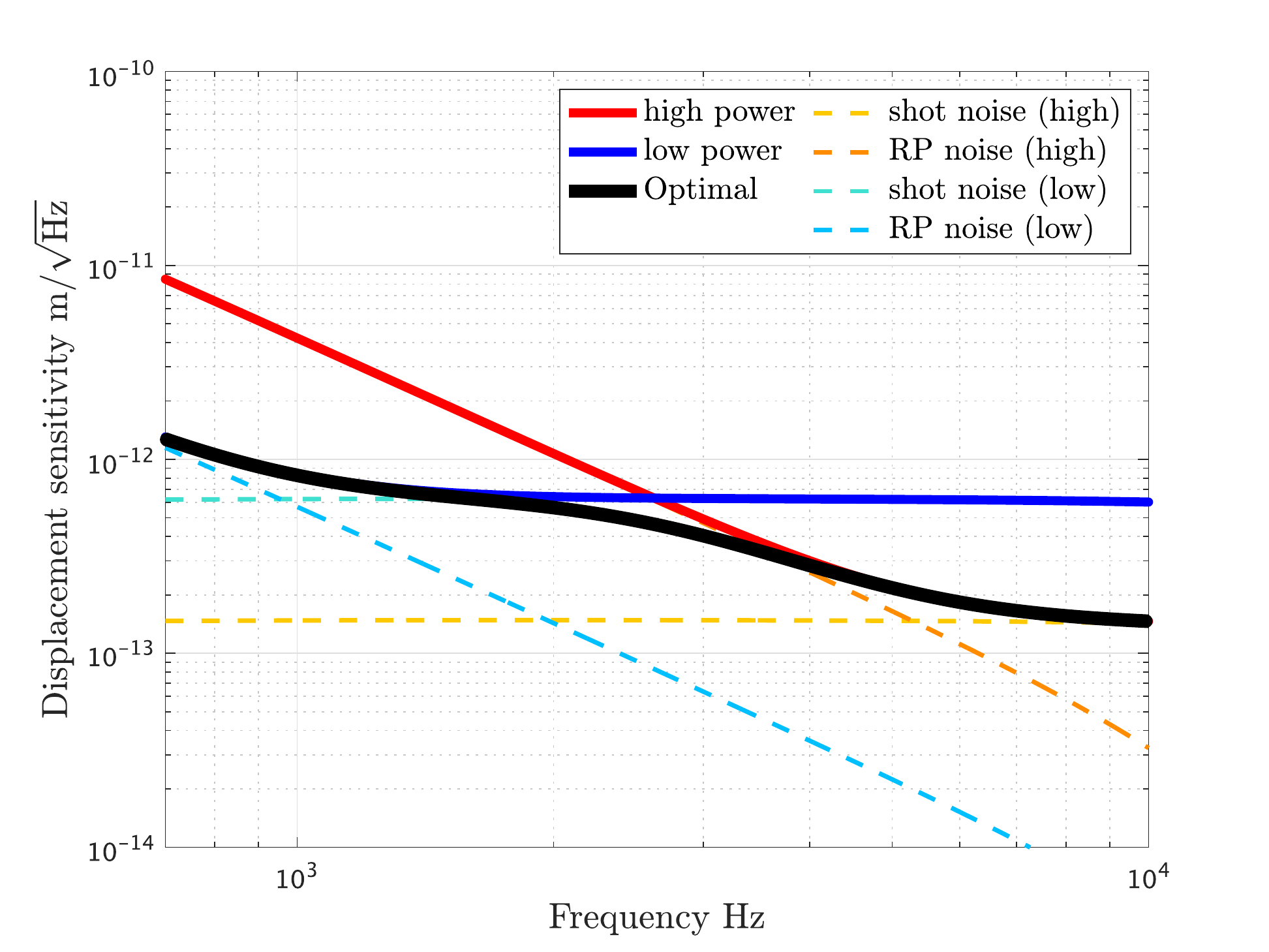}
      \label{fig:7-3}
    }%
    \caption{Measured sensitivities compared with simulated sensitivities. (a) includes the measured sensitivities in the three cases, the modeled quantum noise components in the single cavity cases, and excess background noise of the cavities. The three main curves in (a) are plotted alone in (b). (c) shows simulated sensitivities to compare the experimental results with (b). (a) and (c) share a color scheme for the same noise components.}
    \label{fig:7}
  \end{figure*}

Figure \ref{fig:7-1} displays measured sensitivities in the three cases with the excess background noise of the cavities. The figure includes the noise components in the single cavity cases. The curves of the three cases to verify the essence of the principle are excerpted in Fig.~\ref{fig:7-2}. All the sensitivities and the noise components in Fig.~\ref{fig:7-1} are reproduced in Fig.~\ref{fig:7-3} by simulation. Comparing Fig.~\ref{fig:7-2} with Fig.~\ref{fig:7-3}, we experimentally support the essence of the principle: the optimal sensitivity following the best of the sensitivities in the high/low power single cavity cases can be derived from the combined signal in Eq.~\eqref{eq:3-4} with the square completion method. We also obtain the feature that the optimal sensitivity is better than the sensitivities in the single cavity cases at the cross-over frequency between the two sensitivities. These results successfully verify the principle of the optimization method.

\begin{figure}[htbp]
    \centering
      \includegraphics[keepaspectratio,clip,width=8.0cm]{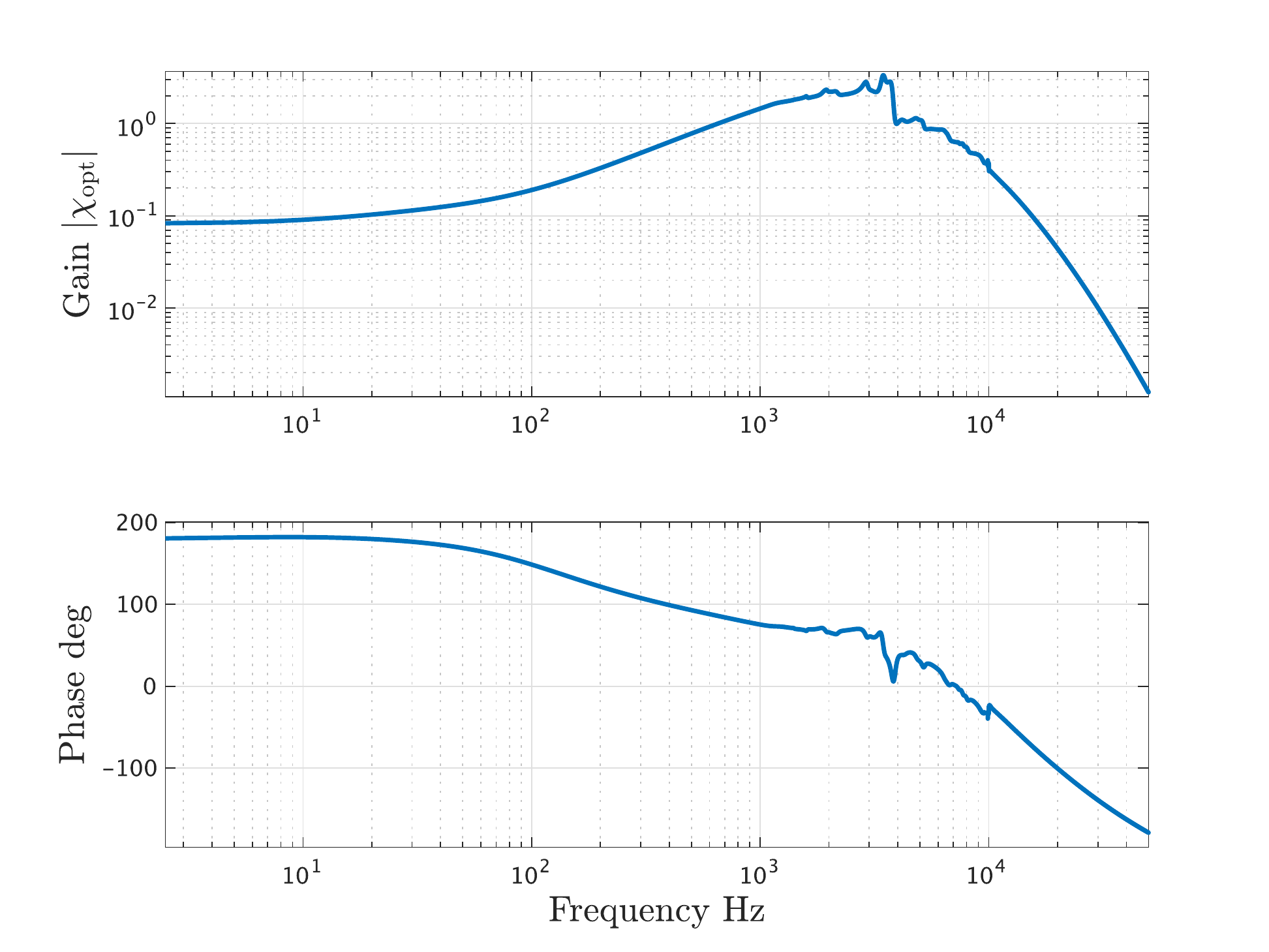}
    \caption{Frequency dependence of the optimized combination coefficient $\chi_\mathrm{opt}$. The mechanical resonances of the mirror-holding structures appear in $\chi_\mathrm{opt}$ at the observation frequency range from 700~Hz to 10~kHz.}
    \label{fig:8}
  \end{figure}

\begin{figure}[htbp]
    \centering
      \includegraphics[keepaspectratio,clip,width=8.0cm]{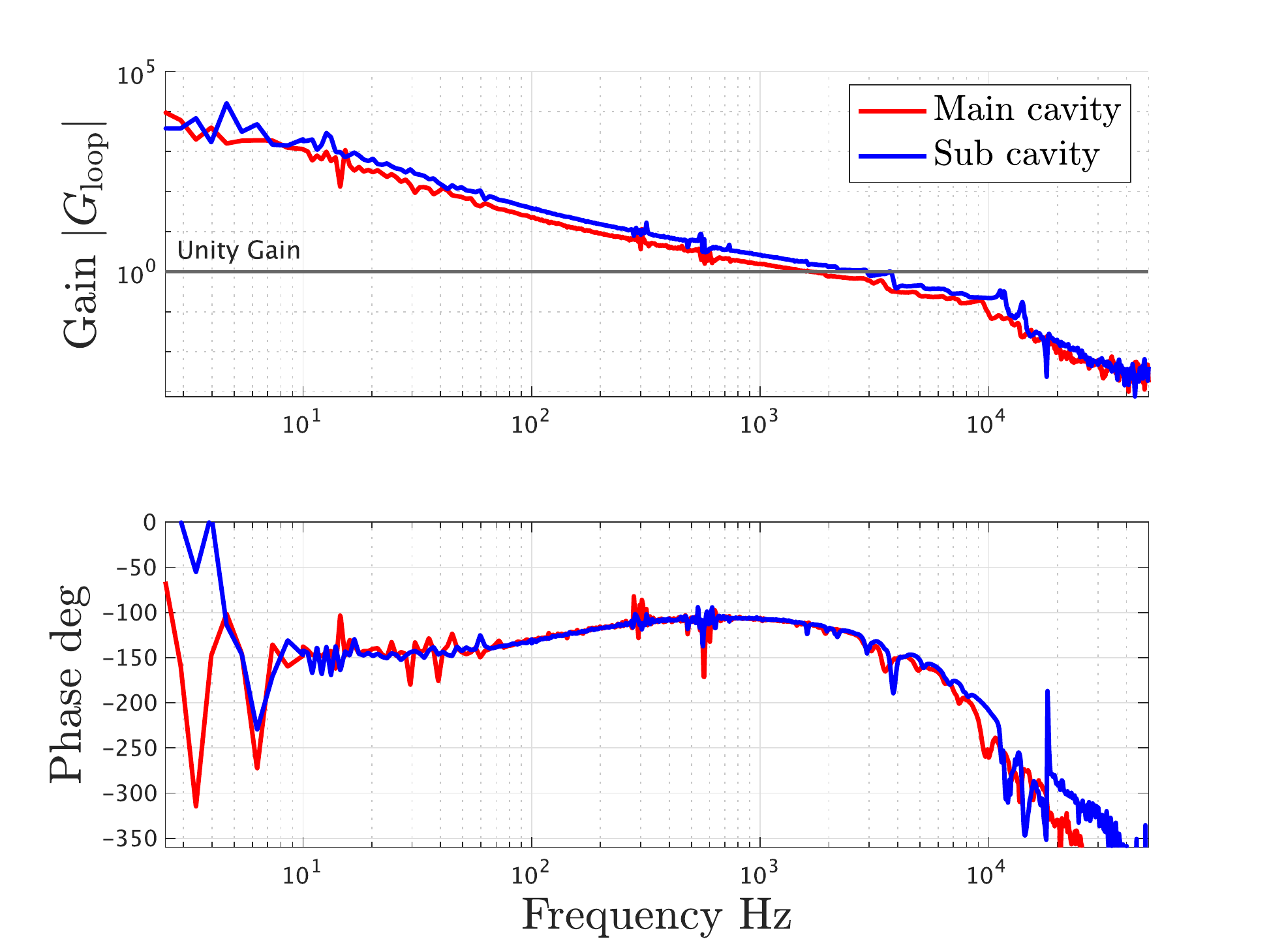}
    \caption{Measured loop gain for the two cavities. The blue/red curves show the loop gain for the main/sub- cavity, respectively.}
    \label{fig:9}
  \end{figure}

Now, we discuss the mechanism of the sensitivity optimization method for the combined signal in Eq.~\eqref{eq:3-4}. The optimal combined signal is obtained by the optimal combination coefficient $\chi_\mathrm{opt}$, which is related to the given loop gain, as discussed in Sec.~\ref{sec:2}. Figure \ref{fig:8} plots the frequency dependence of $\chi_\mathrm{opt}$ in this experiment. We see two features in the figure: $\chi_\mathrm{opt}$ approaches a negative constant value at lower frequencies, and the gain of the $\chi_\mathrm{opt}$ decreases toward zero at high frequencies. Figure \ref{fig:9} shows the measured loop gain in this experiment: the loop gain is high at low frequencies, while the gain continuously decreases at higher frequencies.

Based on the measured loop gain, we can describe the features of $\chi_\mathrm{opt}$ as follows. At the RP-dominated low frequencies, the loop gain has a constant value higher than unity. As mentioned in Sec.~\ref{sec:2}, since the feedback signal suppresses the RP noise of the main cavity but imposes that of the sub-cavity on the displacement of the shared mirror, the RP noise of the sub-cavity contributes more than that of the main cavity to $V_\mathrm{Main}$, and vice versa for $V_\mathrm{Sub}$. The combined signal is thus optimized by utilizing $V_\mathrm{Sub}$ containing the RP noise derived from the main cavity to remove this unnecessary RP noise derived from the main cavity in $V_\mathrm{Main}$
  \begin{equation}
      \tilde{V}_\mathrm{opt} \approx \tilde{V}_\mathrm{Main} - |\chi_\mathrm{opt}| \tilde{V}_\mathrm{Sub}~.
    \label{eq:4-1}
  \end{equation}
At the shot-noise dominated high frequencies, the loop gain decreasing to zero means that the two cavities act as two independent cavities. Since the optimized sensitivity at the high frequencies follows the sensitivity of the high laser power, the combined signal is then optimized by ignoring the error signal $V_\mathrm{Sub}$:
  \begin{equation}
      \tilde{V}_\mathrm{opt} \approx \tilde{V}_\mathrm{Main} + 0 \cdot \tilde{V}_\mathrm{Sub}~.
    \label{eq:4-2}
  \end{equation}
This successfully confirms the principle of the optimization method.

\section{Conclusion}
\label{sec:5}

We have experimentally verified the principle of optimizing the sensitivity for the combined signal by the square completion method. This principle is the first of the three principles which is utilized in the optical-spring quantum locking technique. The experiment employed a tabletop setup under reasonably simplified conditions. The quantum noise was simulated by classical noise. The combined signal from the two cavities is optimized by the square completion method as this provides the optimal combination coefficient $\chi_\mathrm{opt}$ which minimizes the value of the arbitrary quadratic function at every frequency. In our next step, we will experimentally verify the second of the three principles: the optimization method with homodyne detection can beat the SQL of the quantum noise, which will also be simulated by classical noise.

\begin{acknowledgments}
  We would like to thank Dhruva Ganapathy for the English editing. This work was supported by JSPS KAKENHI, Grant Number JP19H01924 and JP22H01247. This work was also supprted by Murata Science Foundation.
\end{acknowledgments}

\newpage
\bibliography{QL_Step1}

\providecommand{\noopsort}[1]{}\providecommand{\singleletter}[1]{#1}%
\begin{thebibliography}{36}%
\makeatletter
\providecommand \@ifxundefined [1]{%
 \@ifx{#1\undefined}
}%
\providecommand \@ifnum [1]{%
 \ifnum #1\expandafter \@firstoftwo
 \else \expandafter \@secondoftwo
 \fi
}%
\providecommand \@ifx [1]{%
 \ifx #1\expandafter \@firstoftwo
 \else \expandafter \@secondoftwo
 \fi
}%
\providecommand \natexlab [1]{#1}%
\providecommand \enquote  [1]{``#1''}%
\providecommand \bibnamefont  [1]{#1}%
\providecommand \bibfnamefont [1]{#1}%
\providecommand \citenamefont [1]{#1}%
\providecommand \href@noop [0]{\@secondoftwo}%
\providecommand \href [0]{\begingroup \@sanitize@url \@href}%
\providecommand \@href[1]{\@@startlink{#1}\@@href}%
\providecommand \@@href[1]{\endgroup#1\@@endlink}%
\providecommand \@sanitize@url [0]{\catcode `\\12\catcode `\$12\catcode
  `\&12\catcode `\#12\catcode `\^12\catcode `\_12\catcode `\%12\relax}%
\providecommand \@@startlink[1]{}%
\providecommand \@@endlink[0]{}%
\providecommand \url  [0]{\begingroup\@sanitize@url \@url }%
\providecommand \@url [1]{\endgroup\@href {#1}{\urlprefix }}%
\providecommand \urlprefix  [0]{URL }%
\providecommand \Eprint [0]{\href }%
\providecommand \doibase [0]{https://doi.org/}%
\providecommand \selectlanguage [0]{\@gobble}%
\providecommand \bibinfo  [0]{\@secondoftwo}%
\providecommand \bibfield  [0]{\@secondoftwo}%
\providecommand \translation [1]{[#1]}%
\providecommand \BibitemOpen [0]{}%
\providecommand \bibitemStop [0]{}%
\providecommand \bibitemNoStop [0]{.\EOS\space}%
\providecommand \EOS [0]{\spacefactor3000\relax}%
\providecommand \BibitemShut  [1]{\csname bibitem#1\endcsname}%
\let\auto@bib@innerbib\@empty
\bibitem [{\citenamefont {{R.~Abbott, $et~al.$}}(2021)}]{Abbott021053}%
  \BibitemOpen
  \bibfield  {author} {\bibinfo {author} {\bibnamefont {{R.~Abbott,
  $et~al.$}}},\ }\bibfield  {title} {\bibinfo {title} {{GWTC-2: Compact Binary
  Coalescences Observed by LIGO and Virgo during the First Half of the Third
  Observing Run}},\ }\href@noop {} {\bibfield  {journal} {\bibinfo  {journal}
  {Phys.\ Rev.\ X}\ }\textbf {\bibinfo {volume} {11}},\ \bibinfo {pages}
  {021053} (\bibinfo {year} {2021})},\ \bibinfo {note}
  {\href{https://doi.org/10.1103/PhysRevX.11.021053}{DOI:
  10.1103/PhysRevX.11.021053}}\BibitemShut {NoStop}%
\bibitem [{\citenamefont {{R.~Abbott, $et~al.$}}()}]{Abbott03606}%
  \BibitemOpen
  \bibfield  {author} {\bibinfo {author} {\bibnamefont {{R.~Abbott,
  $et~al.$}}},\ }\href@noop {} {\bibinfo {title} {{GWTC-3: Compact Binary
  Coalescences Observed by LIGO and Virgo During the Second Part of the Third
  Observing Run}}},\ \bibinfo {howpublished} {e-print arXiv:2111.03606},\
  \bibinfo {note} {\href{https://arxiv.org/abs/2111.03606v2}{arXiv:
  2111.03606}}\BibitemShut {NoStop}%
\bibitem [{\citenamefont {{J.~Aasi, $et~al.$}}(2015)}]{Aasi074001}%
  \BibitemOpen
  \bibfield  {author} {\bibinfo {author} {\bibnamefont {{J.~Aasi, $et~al.$}}},\
  }\bibfield  {title} {\bibinfo {title} {{Advanced LIGO}},\ }\href@noop {}
  {\bibfield  {journal} {\bibinfo  {journal} {Class. Quant. Grav.}\ }\textbf
  {\bibinfo {volume} {32}},\ \bibinfo {pages} {074001} (\bibinfo {year}
  {2015})},\ \bibinfo {note}
  {\href{https://dx.doi.org/10.1088/0264-9381/32/7/074001}{DOI:
  10.1088/0264-9381/32/7/074001}}\BibitemShut {NoStop}%
\bibitem [{\citenamefont {{F.~Acernese, $et~al.$}}(2015)}]{Acernese024001}%
  \BibitemOpen
  \bibfield  {author} {\bibinfo {author} {\bibnamefont {{F.~Acernese,
  $et~al.$}}},\ }\bibfield  {title} {\bibinfo {title} {{Advanced Virgo: a
  second-generation interferometric gravitational wave detector}},\ }\href@noop
  {} {\bibfield  {journal} {\bibinfo  {journal} {Class. Quant. Grav.}\ }\textbf
  {\bibinfo {volume} {32}},\ \bibinfo {pages} {024001} (\bibinfo {year}
  {2015})},\ \bibinfo {note} {\href
  {https://dx.doi.org/10.1088/0264-9381/32/2/024001}{DOI:
  10.1088/0264-9381/32/2/024001}}\BibitemShut {NoStop}%
\bibitem [{\citenamefont {{K.~Somiya}}(2012)}]{Somiya124007}%
  \BibitemOpen
  \bibfield  {author} {\bibinfo {author} {\bibnamefont {{K.~Somiya}}},\
  }\bibfield  {title} {\bibinfo {title} {{Detector configuration of KAGRA–the
  Japanese cryogenic gravitational-wave detector}},\ }\href@noop {} {\bibfield
  {journal} {\bibinfo  {journal} {Class. Quant. Grav.}\ }\textbf {\bibinfo
  {volume} {29}},\ \bibinfo {pages} {124007} (\bibinfo {year} {2012})},\
  \bibinfo {note} {\href
  {https://dx.doi.org/10.1088/0264-9381/29/12/124007}{DOI:
  10.1088/0264-9381/29/12/124007}}\BibitemShut {NoStop}%
\bibitem [{\citenamefont {{M.~Punturo, $et~al.$}}(2010)}]{Punturo084007}%
  \BibitemOpen
  \bibfield  {author} {\bibinfo {author} {\bibnamefont {{M.~Punturo,
  $et~al.$}}},\ }\bibfield  {title} {\bibinfo {title} {{The third generation of
  gravitational wave observatories and their science reach}},\ }\href@noop {}
  {\bibfield  {journal} {\bibinfo  {journal} {Class. Quant. Grav.}\ }\textbf
  {\bibinfo {volume} {27}},\ \bibinfo {pages} {084007} (\bibinfo {year}
  {2010})},\ \bibinfo {note} {\href
  {https://dx.doi.org/10.1088/0264-9381/27/8/084007}{DOI:
  10.1088/0264-9381/27/8/084007}}\BibitemShut {NoStop}%
\bibitem [{\citenamefont {{S.~Dwyer, $et~al.$}}(2015)}]{Dwyer082001}%
  \BibitemOpen
  \bibfield  {author} {\bibinfo {author} {\bibnamefont {{S.~Dwyer,
  $et~al.$}}},\ }\bibfield  {title} {\bibinfo {title} {{Gravitational wave
  detector with cosmological reach}},\ }\href@noop {} {\bibfield  {journal}
  {\bibinfo  {journal} {Phys. Rev. D}\ }\textbf {\bibinfo {volume} {91}},\
  \bibinfo {pages} {082001} (\bibinfo {year} {2015})},\ \bibinfo {note} {\href
  {https://doi.org/10.1103/PhysRevD.91.082001}{DOI:
  10.1103/PhysRevD.91.082001}}\BibitemShut {NoStop}%
\bibitem [{\citenamefont {{Bin~Wu, $et~al.$}}(2022)}]{Bin042007}%
  \BibitemOpen
  \bibfield  {author} {\bibinfo {author} {\bibnamefont {{Bin~Wu, $et~al.$}}},\
  }\bibfield  {title} {\bibinfo {title} {{Conceptual design and science cases
  of a juggled interferometer for gravitational wave detection}},\ }\href@noop
  {} {\bibfield  {journal} {\bibinfo  {journal} {Phys. Rev. D.}\ }\textbf
  {\bibinfo {volume} {106}},\ \bibinfo {pages} {042007} (\bibinfo {year}
  {2022})},\ \bibinfo {note} {\href
  {https://doi.org/10.1103/PhysRevD.106.042007}{DOI:
  10.1103/PhysRevD.106.042007}}\BibitemShut {NoStop}%
\bibitem [{\citenamefont {{S.~Kawamura and Y.~Chen}}(2004)}]{Kawamura211103}%
  \BibitemOpen
  \bibfield  {author} {\bibinfo {author} {\bibnamefont {{S.~Kawamura and
  Y.~Chen}}},\ }\bibfield  {title} {\bibinfo {title} {{Displacement-Noise-Free
  Gravitational-Wave Detection}},\ }\href@noop {} {\bibfield  {journal}
  {\bibinfo  {journal} {Phys. Rev. Lett.}\ }\textbf {\bibinfo {volume} {93}},\
  \bibinfo {pages} {211103} (\bibinfo {year} {2004})},\ \bibinfo {note} {\href
  {https://doi.org/10.1103/PhysRevLett.93.211103}{DOI:
  10.1103/PhysRevLett.93.211103}}\BibitemShut {NoStop}%
\bibitem [{\citenamefont {{S.~Iwaguchi, $et~al.$}}(2022)}]{Iwaguchi128150}%
  \BibitemOpen
  \bibfield  {author} {\bibinfo {author} {\bibnamefont {{S.~Iwaguchi,
  $et~al.$}}},\ }\bibfield  {title} {\bibinfo {title} {{Displacement-noise-free
  neutron interferometer for gravitational wave detection using a single
  Mach-Zehnder configuration}},\ }\href@noop {} {\bibfield  {journal} {\bibinfo
   {journal} {Phys. Lett. A}\ }\textbf {\bibinfo {volume} {441}},\ \bibinfo
  {pages} {128150} (\bibinfo {year} {2022})},\ \bibinfo {note} {\href
  {https://doi.org/10.1016/j.physleta.2022.128150}{DOI:
  10.1016/j.physleta.2022.128150}}\BibitemShut {NoStop}%
\bibitem [{\citenamefont {{K.~Danzmann, $et~al.$}}()}]{Danzmann20170120}%
  \BibitemOpen
  \bibfield  {author} {\bibinfo {author} {\bibnamefont {{K.~Danzmann,
  $et~al.$}}},\ }\bibfield  {title} {\bibinfo {title} {{LISA Laser
  Interferometer Space Antenna.}},\ }\href@noop {} {\ }\bibinfo {note} {\href
  {https://www.elisascience.org/files/publications/LISA\_L3\_20170120.pdf}{URL:
  https://www.elisascience.org/
  files/publications/LISA\_L3\_20170120.pdf}}\BibitemShut {NoStop}%
\bibitem [{\citenamefont {{J.~Crowder and
  N.~J.~Cornish}}(2005)}]{Crowder083005}%
  \BibitemOpen
  \bibfield  {author} {\bibinfo {author} {\bibnamefont {{J.~Crowder and
  N.~J.~Cornish}}},\ }\bibfield  {title} {\bibinfo {title} {{Beyond LISA:
  Exploring future gravitational wave missions}},\ }\href@noop {} {\bibfield
  {journal} {\bibinfo  {journal} {Phys. Rev. D}\ }\textbf {\bibinfo {volume}
  {72}},\ \bibinfo {pages} {083005} (\bibinfo {year} {2005})},\ \bibinfo {note}
  {\href {https://doi.org/10.1103/PhysRevD.72.083005}{DOI:
  10.1103/PhysRevD.72.083005}}\BibitemShut {NoStop}%
\bibitem [{\citenamefont {{N.~Seto, $et~al.$}}(2001)}]{Seto221103}%
  \BibitemOpen
  \bibfield  {author} {\bibinfo {author} {\bibnamefont {{N.~Seto, $et~al.$}}},\
  }\bibfield  {title} {\bibinfo {title} {{Possibility of Direct Measurement of
  the Acceleration of the Universe Using 0.1~Hz Band Laser Interferometer
  Gravitational Wave Antenna in Space}},\ }\href@noop {} {\bibfield  {journal}
  {\bibinfo  {journal} {Phys. Rev. Lett.}\ }\textbf {\bibinfo {volume} {87}},\
  \bibinfo {pages} {221103} (\bibinfo {year} {2001})},\ \bibinfo {note} {\href
  {https://doi.org/10.1103/PhysRevLett.87.221103}{DOI:
  10.1103/PhysRevLett.87.221103}}\BibitemShut {NoStop}%
\bibitem [{\citenamefont {{S. Kawamura, $et~al.$}}(2019)}]{Kawamura1845001}%
  \BibitemOpen
  \bibfield  {author} {\bibinfo {author} {\bibnamefont {{S. Kawamura,
  $et~al.$}}},\ }\bibfield  {title} {\bibinfo {title} {{Space
  gravitational-wave antennas DECIGO and B-DECIGO}},\ }\href@noop {} {\bibfield
   {journal} {\bibinfo  {journal} {Int. J. Mod. Phys. D}\ }\textbf {\bibinfo
  {volume} {28}},\ \bibinfo {pages} {1845001} (\bibinfo {year} {2019})},\
  \bibinfo {note} {\href {https://doi.org/10.1142/S0218271818450013}{DOI:
  10.1142/S0218271818450013}}\BibitemShut {NoStop}%
\bibitem [{\citenamefont {{M.~Maggiore}}(2000)}]{Maggiore283367}%
  \BibitemOpen
  \bibfield  {author} {\bibinfo {author} {\bibnamefont {{M.~Maggiore}}},\
  }\bibfield  {title} {\bibinfo {title} {{Gravitational wave experiments and
  early universe cosmology}},\ }\href@noop {} {\bibfield  {journal} {\bibinfo
  {journal} {Phys. Rep.}\ }\textbf {\bibinfo {volume} {331}},\ \bibinfo {pages}
  {283} (\bibinfo {year} {2000})},\ \bibinfo {note} {\href
  {https://doi.org/10.1016/S0370-1573(99)00102-7}{DOI:
  10.1016/S0370-1573(99)00102-7}}\BibitemShut {NoStop}%
\bibitem [{\citenamefont {{C.~M.~Caves and
  B.~L.~Schumaker}}(1985)}]{Caves3068}%
  \BibitemOpen
  \bibfield  {author} {\bibinfo {author} {\bibnamefont {{C.~M.~Caves and
  B.~L.~Schumaker}}},\ }\bibfield  {title} {\bibinfo {title} {{New formalism
  for two-photon quantum optics. I. Quadrature phases and squeezed states}},\
  }\href@noop {} {\bibfield  {journal} {\bibinfo  {journal} {Phys. Rev. A}\
  }\textbf {\bibinfo {volume} {31}},\ \bibinfo {pages} {3068} (\bibinfo {year}
  {1985})},\ \bibinfo {note} {\href
  {https://doi.org/10.1103/PhysRevA.31.3068}{DOI:
  10.1103/PhysRevA.31.3068}}\BibitemShut {NoStop}%
\bibitem [{\citenamefont {{B.~L.~Schumaker and
  C.~M.~Caves}}(1985)}]{Schumaker3093}%
  \BibitemOpen
  \bibfield  {author} {\bibinfo {author} {\bibnamefont {{B.~L.~Schumaker and
  C.~M.~Caves}}},\ }\bibfield  {title} {\bibinfo {title} {{New formalism for
  two-photon quantum optics. II. Mathematical foundation and compact
  notation}},\ }\href@noop {} {\bibfield  {journal} {\bibinfo  {journal} {Phys.
  Rev. A}\ }\textbf {\bibinfo {volume} {31}},\ \bibinfo {pages} {3093}
  (\bibinfo {year} {1985})},\ \bibinfo {note} {\href
  {https://doi.org/10.1103/PhysRevA.31.3093}{DOI:
  10.1103/PhysRevA.31.3093}}\BibitemShut {NoStop}%
\bibitem [{\citenamefont {{Y.~Akrami, $et~al.$}}(2020)}]{Akrami641A10}%
  \BibitemOpen
  \bibfield  {author} {\bibinfo {author} {\bibnamefont {{Y.~Akrami,
  $et~al.$}}},\ }\bibfield  {title} {\bibinfo {title} {{Planck 2018 results X.
  Constraints on inflation}},\ }\href@noop {} {\bibfield  {journal} {\bibinfo
  {journal} {{A \& A}}\ }\textbf {\bibinfo {volume} {641}},\ \bibinfo {pages}
  {A10} (\bibinfo {year} {2020})},\ \bibinfo {note} {\href
  {https://doi.org/10.1051/0004-6361/201833887}{DOI:
  10.1051/0004-6361/201833887}}\BibitemShut {NoStop}%
\bibitem [{\citenamefont {{S.~Kuroyanagi, $et~al.$}}(2014)}]{Kuroyanagi063513}%
  \BibitemOpen
  \bibfield  {author} {\bibinfo {author} {\bibnamefont {{S.~Kuroyanagi,
  $et~al.$}}},\ }\bibfield  {title} {\bibinfo {title} {{Implications of the
  $B$-mode polarization measurement for direct detection of inflationary
  gravitational waves}},\ }\href@noop {} {\bibfield  {journal} {\bibinfo
  {journal} {Phys. Rev. D}\ }\textbf {\bibinfo {volume} {90}},\ \bibinfo
  {pages} {063513} (\bibinfo {year} {2014})},\ \bibinfo {note} {\href
  {https://doi.org/10.1103/PhysRevD.90.063513}{DOI:
  10.1103/PhysRevD.90.063513}}\BibitemShut {NoStop}%
\bibitem [{\citenamefont {{S.~Iwaguchi, $et~al.$}}(2021)}]{Iwaguchi9010009}%
  \BibitemOpen
  \bibfield  {author} {\bibinfo {author} {\bibnamefont {{S.~Iwaguchi,
  $et~al.$}}},\ }\bibfield  {title} {\bibinfo {title} {{Quantum Noise in a
  Fabry-Perot Interferometer Including the Influence of Diffraction Loss of
  Light}},\ }\href@noop {} {\bibfield  {journal} {\bibinfo  {journal}
  {Galaxies}\ }\textbf {\bibinfo {volume} {9}},\ \bibinfo {pages} {9} (\bibinfo
  {year} {2021})},\ \bibinfo {note} {\href
  {https://doi.org/10.3390/galaxies9010009}{DOI:
  10.3390/galaxies9010009}}\BibitemShut {NoStop}%
\bibitem [{\citenamefont {{T.~Ishikawa, $et~al.$}}(2021)}]{Ishikawa9010014}%
  \BibitemOpen
  \bibfield  {author} {\bibinfo {author} {\bibnamefont {{T.~Ishikawa,
  $et~al.$}}},\ }\bibfield  {title} {\bibinfo {title} {{Improvement of the
  Target Sensitivity in DECIGO by Optimizing its Parameters for Quantum Noise
  Including the Effect of Diffraction Loss}},\ }\href@noop {} {\bibfield
  {journal} {\bibinfo  {journal} {Galaxies}\ }\textbf {\bibinfo {volume} {9}},\
  \bibinfo {pages} {14} (\bibinfo {year} {2021})},\ \bibinfo {note} {\href
  {https://doi.org/10.3390/galaxies9010014}{DOI:
  10.3390/galaxies9010014}}\BibitemShut {NoStop}%
\bibitem [{\citenamefont {{Y.~Kawasaki, $et~al.$}}(2022)}]{Kawasaki10010025}%
  \BibitemOpen
  \bibfield  {author} {\bibinfo {author} {\bibnamefont {{Y.~Kawasaki,
  $et~al.$}}},\ }\bibfield  {title} {\bibinfo {title} {{Optimization of Design
  Parameters for Gravitational Wave Detector DECIGO Including Fundamental
  Noises}},\ }\href@noop {} {\bibfield  {journal} {\bibinfo  {journal}
  {Galaxies}\ }\textbf {\bibinfo {volume} {10}},\ \bibinfo {pages} {25}
  (\bibinfo {year} {2022})},\ \bibinfo {note} {\href
  {https://doi.org/10.3390/galaxies10010025}{DOI:
  10.3390/galaxies10010025}}\BibitemShut {NoStop}%
\bibitem [{\citenamefont {{H.~J.~Kimble, $et~al.$}}(2001)}]{Kimble022002}%
  \BibitemOpen
  \bibfield  {author} {\bibinfo {author} {\bibnamefont {{H.~J.~Kimble,
  $et~al.$}}},\ }\bibfield  {title} {\bibinfo {title} {{Conversion of
  conventional gravitational-wave interferometers into quantum nondemolition
  interferometers by modifying their input and/or output optics}},\ }\href@noop
  {} {\bibfield  {journal} {\bibinfo  {journal} {Phys. Rev. D}\ }\textbf
  {\bibinfo {volume} {65}},\ \bibinfo {pages} {022002} (\bibinfo {year}
  {2001})},\ \bibinfo {note} {\href
  {https://doi.org/10.1103/PhysRevD.65.022002}{DOI:
  10.1103/PhysRevD.65.022002}}\BibitemShut {NoStop}%
\bibitem [{\citenamefont {{The LIGO Scientific
  Collaboration}}(2011)}]{LIGOCollab2083}%
  \BibitemOpen
  \bibfield  {author} {\bibinfo {author} {\bibnamefont {{The LIGO Scientific
  Collaboration}}},\ }\bibfield  {title} {\bibinfo {title} {{A gravitational
  wave observatory operating beyond the quantum shot-noise limit}},\
  }\href@noop {} {\bibfield  {journal} {\bibinfo  {journal} {Nat. Phys.}\
  }\textbf {\bibinfo {volume} {7}},\ \bibinfo {pages} {962} (\bibinfo {year}
  {2011})},\ \bibinfo {note} {\href {https://doi.org/10.1038/nphys2083}{DOI:
  10.1038/nphys2083}}\BibitemShut {NoStop}%
\bibitem [{\citenamefont {{J.~Aasi, $et~al.$}}(2013)}]{Aasi177}%
  \BibitemOpen
  \bibfield  {author} {\bibinfo {author} {\bibnamefont {{J.~Aasi, $et~al.$}}},\
  }\bibfield  {title} {\bibinfo {title} {{Enhanced sensitivity of the LIGO
  gravitational wave detector by using squeezed states of light}},\ }\href@noop
  {} {\bibfield  {journal} {\bibinfo  {journal} {Nat. Photonics}\ }\textbf
  {\bibinfo {volume} {7}},\ \bibinfo {pages} {613} (\bibinfo {year} {2013})},\
  \bibinfo {note} {\href {https://doi.org/10.1038/nphoton.2013.177}{DOI:
  10.1038/nphoton.2013.177}}\BibitemShut {NoStop}%
\bibitem [{\citenamefont {{F.~Acernese, $et~al.$}}(2019)}]{Acernese321108}%
  \BibitemOpen
  \bibfield  {author} {\bibinfo {author} {\bibnamefont {{F.~Acernese,
  $et~al.$}}},\ }\bibfield  {title} {\bibinfo {title} {{Increasing the
  astrophysical reach of the advanced Virgo detector via the application of
  squeezed vacuum states of light}},\ }\href@noop {} {\bibfield  {journal}
  {\bibinfo  {journal} {Phys. Rev. Lett.}\ }\textbf {\bibinfo {volume} {123}},\
  \bibinfo {pages} {321108} (\bibinfo {year} {2019})},\ \bibinfo {note} {\href
  {https://doi.org/10.1103/PhysRevLett.123.231108}{DOI:
  10.1103/PhysRevLett.123.231108}}\BibitemShut {NoStop}%
\bibitem [{\citenamefont {{H.~Abu-Safia, $et~al.$}}(1994)}]{Safia003805}%
  \BibitemOpen
  \bibfield  {author} {\bibinfo {author} {\bibnamefont {{H.~Abu-Safia,
  $et~al.$}}},\ }\bibfield  {title} {\bibinfo {title} {{Transmission of a
  Gaussian beam through a Fabry-Perot interferometer}},\ }\href@noop {}
  {\bibfield  {journal} {\bibinfo  {journal} {Appl. Opt.}\ }\textbf {\bibinfo
  {volume} {33}},\ \bibinfo {pages} {3805} (\bibinfo {year} {1994})},\ \bibinfo
  {note} {\href {https://doi.org/10.1364/AO.33.003805}{DOI:
  10.1364/AO.33.003805}}\BibitemShut {NoStop}%
\bibitem [{\citenamefont {{R.~Yamada, $et~al.$}}(2020)}]{Yamada126626}%
  \BibitemOpen
  \bibfield  {author} {\bibinfo {author} {\bibnamefont {{R.~Yamada,
  $et~al.$}}},\ }\bibfield  {title} {\bibinfo {title} {{Optimization of quantum
  noise by completing the square of multiple interferometer outputs in quantum
  locking for gravitational wave detectors}},\ }\href@noop {} {\bibfield
  {journal} {\bibinfo  {journal} {Phys. Lett. A}\ }\textbf {\bibinfo {volume}
  {384}},\ \bibinfo {pages} {126626} (\bibinfo {year} {2020})},\ \bibinfo
  {note} {\href {https://doi.org/10.1016/j.physleta.2020.126626}{DOI:
  j.physleta.2020.126626}}\BibitemShut {NoStop}%
\bibitem [{\citenamefont {{R.~Yamada, $et~al.$}}(2021)}]{Yamada127365}%
  \BibitemOpen
  \bibfield  {author} {\bibinfo {author} {\bibnamefont {{R.~Yamada,
  $et~al.$}}},\ }\bibfield  {title} {\bibinfo {title} {{Reduction of quantum
  noise using the quantum locking with an optical spring for gravitational wave
  detectors}},\ }\href@noop {} {\bibfield  {journal} {\bibinfo  {journal}
  {Phys. Lett. A}\ }\textbf {\bibinfo {volume} {402}},\ \bibinfo {pages}
  {127365} (\bibinfo {year} {2021})},\ \bibinfo {note} {\href
  {https://doi.org/10.1016/j.physleta.2021.127365}{DOI:
  10.1016/j.physleta.2021.127365}}\BibitemShut {NoStop}%
\bibitem [{\citenamefont {{J.~M.~Courty, $et~al.$}}(2003)}]{Courty083601}%
  \BibitemOpen
  \bibfield  {author} {\bibinfo {author} {\bibnamefont {{J.~M.~Courty,
  $et~al.$}}},\ }\bibfield  {title} {\bibinfo {title} {{Quantum Locking of
  Mirrors in Interferometers}},\ }\href@noop {} {\bibfield  {journal} {\bibinfo
   {journal} {Phys. Rev. Lett.}\ }\textbf {\bibinfo {volume} {90}},\ \bibinfo
  {pages} {083601} (\bibinfo {year} {2003})},\ \bibinfo {note} {\href
  {https://doi.org/10.1103/PhysRevLett.90.083601}{DOI:
  10.1103/PhysRevLett.90.083601}}\BibitemShut {NoStop}%
\bibitem [{\citenamefont {{A.~Heidmann, $et~al.$}}(2004)}]{HeidmannS684}%
  \BibitemOpen
  \bibfield  {author} {\bibinfo {author} {\bibnamefont {{A.~Heidmann,
  $et~al.$}}},\ }\bibfield  {title} {\bibinfo {title} {{Beating quantum limits
  in interferometers with quantum locking of mirrors}},\ }\href@noop {}
  {\bibfield  {journal} {\bibinfo  {journal} {J. Opt. B}\ }\textbf {\bibinfo
  {volume} {6}},\ \bibinfo {pages} {S684} (\bibinfo {year} {2004})},\ \bibinfo
  {note} {\href {https://doi.org/10.1088/1464-4266/6/8/009}{DOI:
  10.1088/1464-4266/6/8/009}}\BibitemShut {NoStop}%
\bibitem [{\citenamefont {{O.~Arcizet, $et~al.$}}(2006)}]{Arcizet033819}%
  \BibitemOpen
  \bibfield  {author} {\bibinfo {author} {\bibnamefont {{O.~Arcizet,
  $et~al.$}}},\ }\bibfield  {title} {\bibinfo {title} {{Beating quantum limits
  in an optomechanical sensor by cavity detuning}},\ }\href@noop {} {\bibfield
  {journal} {\bibinfo  {journal} {Phys. Rev. A}\ }\textbf {\bibinfo {volume}
  {73}},\ \bibinfo {pages} {033819} (\bibinfo {year} {2006})},\ \bibinfo {note}
  {\href {https://doi.org/10.1103/PhysRevA.73.033819}{DOI:
  10.1103/PhysRevA.73.033819}}\BibitemShut {NoStop}%
\bibitem [{\citenamefont {{P.~Verlot, $et~al.$}}(2010)}]{Verlot133602}%
  \BibitemOpen
  \bibfield  {author} {\bibinfo {author} {\bibnamefont {{P.~Verlot,
  $et~al.$}}},\ }\bibfield  {title} {\bibinfo {title} {{Back-action
  amplification and quantum limits in optomechanical measurements}},\
  }\href@noop {} {\bibfield  {journal} {\bibinfo  {journal} {Phys. Rev. Lett.}\
  }\textbf {\bibinfo {volume} {104}},\ \bibinfo {pages} {133602} (\bibinfo
  {year} {2010})},\ \bibinfo {note} {\href
  {https://doi.org/10.1103/PhysRevLett.104.133602}{DOI:
  10.1103/PhysRevLett.104.133602}}\BibitemShut {NoStop}%
\bibitem [{\citenamefont {{C.~M.~Caves}}(1980)}]{Caves75}%
  \BibitemOpen
  \bibfield  {author} {\bibinfo {author} {\bibnamefont {{C.~M.~Caves}}},\
  }\bibfield  {title} {\bibinfo {title} {{Quantum-Mechanical Radiation-Pressure
  Fluctuations in an Interferometer}},\ }\href@noop {} {\bibfield  {journal}
  {\bibinfo  {journal} {Phys. Rev. Lett.}\ }\textbf {\bibinfo {volume} {45}},\
  \bibinfo {pages} {75} (\bibinfo {year} {1980})},\ \bibinfo {note} {\href
  {https://doi.org/10.1103/PhysRevLett.45.75}{DOI:
  10.1103/PhysRevLett.45.75}}\BibitemShut {NoStop}%
\bibitem [{\citenamefont {{C.~M.~Caves}}(1981)}]{Caves1693}%
  \BibitemOpen
  \bibfield  {author} {\bibinfo {author} {\bibnamefont {{C.~M.~Caves}}},\
  }\bibfield  {title} {\bibinfo {title} {{Quantum-mechanical noise in an
  interferometer}},\ }\href@noop {} {\bibfield  {journal} {\bibinfo  {journal}
  {Phys. Rev. D}\ }\textbf {\bibinfo {volume} {23}},\ \bibinfo {pages} {1693}
  (\bibinfo {year} {1981})},\ \bibinfo {note} {\href
  {https://doi.org/10.1103/PhysRevD.23.1693}{DOI:
  10.1103/PhysRevD.23.1693}}\BibitemShut {NoStop}%
\bibitem [{\citenamefont {{R.~W.~P.~Drever, $et~al.$}}(1983)}]{Drever00702605}%
  \BibitemOpen
  \bibfield  {author} {\bibinfo {author} {\bibnamefont {{R.~W.~P.~Drever,
  $et~al.$}}},\ }\bibfield  {title} {\bibinfo {title} {{Laser phase and
  frequency stabilization using an optical resonator}},\ }\href@noop {}
  {\bibfield  {journal} {\bibinfo  {journal} {Applied Physics B}\ }\textbf
  {\bibinfo {volume} {31}},\ \bibinfo {pages} {97} (\bibinfo {year} {1983})},\
  \bibinfo {note} {\href {https://doi.org/10.1007/BF00702605}{DOI:
  10.1007/BF00702605}}\BibitemShut {NoStop}%
\end{thebibliography}%


\providecommand{\noopsort}[1]{}\providecommand{\singleletter}[1]{#1}%
%

\end{document}